\documentclass[12pt,preprint]{aastex}
\usepackage{lscape}

\usepackage{amsmath}

\shorttitle{$z\sim3$ galaxy correlation functions}
\shortauthors{Cooke et al.}

\begin{document}

\title{Survey for Galaxies Associated with $z\sim3$ Damped Lyman
  $\alpha$ Systems II: Galaxy-Absorber Correlation Functions}

\author{Jeff Cooke\altaffilmark{1} \altaffilmark{2}
\& Arthur M. Wolfe\altaffilmark{1}}
\affil{Department of Physics and Center for Astrophysics and
  Space Sciences; University of California, San Diego, mc-0424, La
  Jolla, CA 92093-0424}
\email{cooke@uci.edu}
\email{awolfe@ucsd.edu}

\author{Eric Gawiser\altaffilmark{1}\altaffilmark{3}}
\affil{Department of Astronomy; Yale University, P.O. Box 208101, New
  Haven, CT 06520-8101}
\email{gawiser@astro.yale.edu}

\and

\author{Jason X. Prochaska\altaffilmark{1}}
\affil{UCO-Lick Observatory; University of California, Santa Cruz,
  Santa Cruz, CA 95064}
\email{xavier@ucolick.edu}

\altaffiltext{1}{Visiting Astronomer, W. M. Keck Telescope. The Keck
  Observatory is a joint facility of the University of California, the
  California Institute of Technology, and NASA and was made possible
  by the generous financial support of the W. M. Keck Foundation.}
\altaffiltext{2}{Current address: Center for Cosmology, University
  of California, Irvine, CA 92697} 
\altaffiltext{3}{NSF Astronomy \& Astrophysics Postdoctoral Fellow}

\begin{abstract} 

We use 211 galaxy spectra from our survey for Lyman break galaxies
(LBGs) associated with 11 damped Lyman $\alpha$ systems (DLAs) to
measure the three-dimensional LBG auto-correlation and DLA-LBG
cross-correlation functions with the primary goal of inferring the
mass of DLAs at $z\sim3$.  From every measurement and test in this
work, we find evidence for an overdensity of LBGs near DLAs and find
this overdensity to be very similar to that of LBGs near other LBGs.
Conventional binning of the data while varying both $r_0$ and $\gamma$
parameters of the fiducial model of the correlation function $\xi(r) =
(r/r_0)^{-\gamma}$ resulted in the best fit values and $1 \sigma$
uncertainties of $r_0=2.65\pm 0.48, \gamma=1.55\pm 0.40$ for the LBG
auto-correlation and $r_0=3.32\pm 1.25, \gamma=1.74\pm 0.36$ for
DLA-LBG cross-correlation function.  To circumvent shortcomings found
in binning small datasets, we perform a maximum likelihood analysis
based on Poisson statistics.  The best fit values and $1 \sigma$
confidence levels from this analysis were found to be $r_0 =
2.91^{+1.0}_{-1.0},~\gamma = 1.21^{+0.6}_{-0.3}$ for the LBG
auto-correlation and $r_0 = 2.81^{+1.4}_{-2.0}$, $\gamma =
2.11^{+1.3}_{-1.4}$ for the DLA-LBG cross-correlation function.  We
report a redshift spike of five LBGs with $Delta z = 0.015$ of the
$z=2.936$ DLA in the PSS0808+5215 field and find that the DLA-LBG
clustering signal survives when omitting this field from the analysis.
Using the correlation functions measurements and uncertainties, we
compute the $z\sim3$ LBG galaxy bias $b_{LBG}$ to be $1.5<b_{LBG}<3$
corresponding to an average halo mass of $10^{9.7}< \langle
M_{LBG}\rangle< 10^{11.6}~M_{\odot}$ and the $z\sim3$ DLA galaxy bias
$b_{DLA}$ to be $1.3<b_{DLA}<4$ corresponding to an average halo mass
of $10^{9}< \langle M_{LBG}\rangle< 10^{12}~M_{\odot}$.  Lastly, two
of the six QSOs discovered in this survey were found to lie within
$\Delta z = 0.0125$ of two of the survey DLAs.  We estimate the
probability of this occurring by chance is 1 in 940 and may indicate
a possible relationship between the distribution of QSOs and DLAs at
$z\sim3$.
\end{abstract}

\keywords{galaxies: high redshift --- quasars: absorption lines ---  
galaxies: formation}

\section{Introduction}   

One of the most fundamental measurements of high redshift galaxies is
their spatial distribution, largely because this information can be
used to infer their average dark matter halo mass.  Standard CDM
cosmology details the process in which galaxies formed by the
gravitational collapse of primordial dark matter density fluctuations.
In the early universe, low-mass fluctuations typically had density
contrasts high enough to collapse against the Hubble flow and formed
uniformly throughout space whereas high-mass fluctuations with high
density contrasts were rare.  However, the superposition of density
fluctuations resulted in the collapse of an excess number of high-mass
fluctuations clustered near the peaks of underlying mass overdensities
that sufficiently enhanced their density contrast.  It is in this
context that the measurement of the clustering of galaxies infers the
underlying dark matter halo mass of these systems.

Several surveys over the last decade have been used to infer the dark
matter mass of high redshift galaxy populations by their angular
clustering [e.g., \citet{d00,d04,mc01,mc04,porc02,foucaud03}].  One
population, the Lyman break galaxies (LBGs), are identified
photometrically by the decrement (break) in their continua shortward
of the Lyman limit caused by absorption from optically thick
intervening systems in the line-of-sight.  The spectroscopic survey of
\citet{s98} provided sufficient spectra for a measurement of the
three-dimensional distribution of LBGs at $z\sim3$. Analysis of that
data by \citet{a98} showed evidence of significant clustering
corresponding to halo masses of $\sim10^{11-12}M_\odot$.  The faint
emission from LBGs restricts observations to low signal-to-noise ratio
low-resolution spectra using current technology.  As a consequence,
the properties of this magnitude limited sample must be examined
statistically and/or from composite spectra.  In addition, the
selection methods employed to detect LBGs are not sensitive to all
types of galaxies at high redshift and therefore partially incomplete.

Quasar absorption-line systems provide a complementary means to study
high redshift systems.  Their detection is dependent only on the
brightness of the background source and the strength of the
absorption-line features and is not biased by intrinsic magnitudes or
photometric selection criteria.  A subset of this population, the
damped Lyman $\alpha$ systems (DLAs), are defined to have column
densities of N(\textsc{Hi}) $>2\times10^{20}$ atoms cm$^{-1}$
\citep{w86,w05} and contain $\sim80\%$ of the \textsc{Hi} content of
the universe \citep{prochaska05}.  DLAs have column densities high
enough to provide self-shielding against ambient ionizing radiation
and thereby protect large reservoirs of neutral gas.  Several lines of
evidence, including high-resolution analysis of DLA gas kinematics
\citep{pw97,pw98,wp00a,wp00b} and the agreement between the comoving
neutral gas density at $z>2$ and the mass density of visible stars in
local disks \citep{w95}, have fueled the belief that DLAs are capable
of evolving into present-day galaxies like the Milky Way \citep{k96}.

It becomes clear that studies of galaxy formation are incomplete
without understanding the population of proto-galaxies that DLAs
represent.  Although properties such as the gas kinematics and
chemical abundances of DLAs can be studied with high resolution and
are well documented, the mass of these systems has remained unknown.
The sparse distribution of DLAs in QSO sight-lines prohibits the use
of their clustering as a tracer of the underlying dark matter mass.
Instead, the host halo mass of DLAs can be inferred by their
cross-correlation with another known population.  In this work, we
utilize the ubiquitous LBGs at $z\sim3$ for this purpose.  This paper
is the second in a series (\citet{c05}, hereafter Paper I) reporting
the results of a survey of LBGs detected near 11 DLAs at $z\sim3$ with
the primary goal of inferring the mass of DLAs from the DLA-LBG
spatial cross-correlation.

This paper is organized as follows: We briefly review the imaging and
spectroscopic observations in \S\ref{observations}, present spectra of
the objects detected near the DLAs in \S\ref{assoc} and the spectra
of the faint AGN discovered in the survey in \S\ref{AGN}.  We describe
the methodology behind the clustering analysis in \S\ref{methodology}
and present the LBG auto-correlation and DLA-LBG cross-correlation
results by technique, including tests of these results, in
\S\ref{results}.  We discuss the implications on the galaxy bias and
mass of DLAs in \S\ref{mass}, provide a brief discussion in
\S\ref{disc} and a summary of the survey in \S\ref{resultssummary}.
We adopt $\Omega_{M}$=0.3, $\Omega_{\Lambda}$=0.7, $h=100$ cosmology
unless otherwise noted, primarily for direct comparison to the
literature.  All correlation lengths $r_0$ reported in this work are
in $h^{-1}$Mpc comoving.

\section{Survey Design}\label{observations}   

Our first goal was to efficiently select LBGs at $z\sim3$ via their
photometric colors.  To achieve this, we developed a $u'$ BVRI
photometric selection technique, primarily due to filter availability,
that fulfilled the imaging goals and resulted in an efficacy very
similar to previous work.  Our second observational goal was the
accurate acquisition and analysis of a few hundred faint $z\sim3$
galaxy spectra.  This was accomplished using the Keck telescope and
low-resolution multi-object spectroscopy. A detailed description of
the data acquisition, reduction and analysis and comparison to
previous $U_nG\cal{R}$I surveys can be found in Paper I. Here, we
discuss a few relevant points of the observations.

\subsection{The DLA Sample}

The QSO fields targeted in this survey were selected from the subset
that met the following constraints: (1) The QSO was required to
display at least one known DLA in the redshift range where our
$u'$BVRI photometric selection is most effective ($2.6<z<3.4$), (2)
systems within 3000 km s$^{-1}$ from the background QSO were not used
to avoid proximity effects and, (3) a preference was given to fields
that would be observed through the lowest airmass and in the direction
of lowest Galactic reddening to minimize the flux attenuation of
$z\sim3$ sources.  Table~\ref{table:qsofields} lists our survey
fields.  At the time this survey began, only $\sim30$ DLAs met these
criteria.  From this subset, the targets were chosen at random.  The
mean column density of the 11 DLAs in this survey is log
N(\textsc{Hi}) $=20.94$ atoms cm$^{-2}$ with a typical individual
error of $\pm0.1$, whereas the mean DLA column density measured
recently from 197 DLAs taken over the same redshift range from the
SDSS survey \citep{prochaska05} is N(\textsc{Hi}) $=20.84$ atoms
cm$^{-2}$ and a typical error of $\pm0.2$.  Therefore, the DLAs
presented here appear to be a good representation of $z\sim3$ DLAs on
average.

\subsection{Imaging}

We obtained deep $u'$BVRI imaging of nine QSO fields with 11 known
$z\sim3$ DLAs from 2000 April to 2003 November using the
Low-Resolution Imager and Spectrometer \citep{o95} on the 10m Keck I
telescope and the Carnegie Observatories Spectrograph and Multi-object
Imaging Camera \citep{k98} on the 5m Hale telescope at the Palomar
Observatory.  Object placement and image field size are topics that
deserve a brief discussion.  The DLAs were approximately centered in
the images in which seven have a field of view of $\sim6'\times7.5'$
(LRIS) and two have a field of view $\sim9.7'\times9.7'$ (COSMIC).
The usable area of each of the final stacked images was reduced in
both dimensions by $\lesssim 40 "$ from our imposed dithering
sequence.  This resulted in maximum (diagonal) comoving object
separations of $\sim10 h^{-1}$Mpc (LRIS) and $\sim13 h^{-1}$Mpc
(COSMIC) at $z\sim3$ and about half this distance for the DLAs.  The
correlation length of LBGs at $z\sim3$ has been measured to be
$3.96\pm0.29 h^{-1}$Mpc by \citet{a03}, hereafter A03.  As a result,
the angular clustering component in our survey cannot be measured much
beyond $1-2$ correlation lengths, yet the redshift path of
$\sim540h^{-1}$Mpc, where our photometric selection is best described
($2.6<z<3.4$), allows complete analysis of the distribution of the
LBGs in the redshift direction.

\subsubsection{Photometric color selection}

We searched for starforming galaxies at $z\sim3$ with expectations and
spectral profiles described in previous work
\citep{s96,st96,l97,pet00,aes03}].  The $u'$BVRI filters are
well-suited to select these objects via their broadband colors.  It is
important to note that surveys with these expectations, although
efficient in detecting a large number of galaxies at $z\sim3$, are not
sensitive to all galaxies that may exist at high redshift and the
corresponding underlying dark matter they may trace.  For example,
systems with excessive intrinsic extinction or older stellar
populations will not be selected.  In our effort to mimic previous
surveys, we remain partially incomplete to all galaxies at $z\sim3$,
but include a significant number to provide meaningful statistics
toward the goal of cross-correlating LBGs with DLAs. From the complete
list of sources detected in each field, we selected LBG candidates
that met the following color constraints
\begin{equation}\label{uB}
(u'-B)_{AB} > 1.1
\end{equation}
\begin{equation}
(u'-V)_{AB} > 1.6
\end{equation}
\begin{equation}
0.6 < (B-R)_{AB} < 2.1
\end{equation}
\begin{equation}
(V-R)_{AB} < 0.6
\end{equation}
\begin{equation}\label{VI}
(V-I)_{AB} < 0.6,~\textup{and}
\end{equation}
\begin{equation}\label{eq:Rmag}
20.0 < R_{AB} < 25.5
\end{equation}

\noindent and assigned these candidates the highest priority.  Some
candidates were selected using a subset of these constraints and
assigned a lower priority.  An additional set of candidates was
selected by relaxing equations~\ref{uB} --~\ref{eq:Rmag} by 0.2
magnitudes and were assigned the lower respective priority.  This was
done in an effort to include objects that would be overlooked due to
photometric errors arising from statistical and systematic
uncertainties.  All LBG candidates were then slated by priority for
follow-up spectroscopy.

\subsection{Spectroscopy}\label{spectro}

\begin{figure}
\begin{center}                          
\scalebox{0.6}[0.6]{\includegraphics{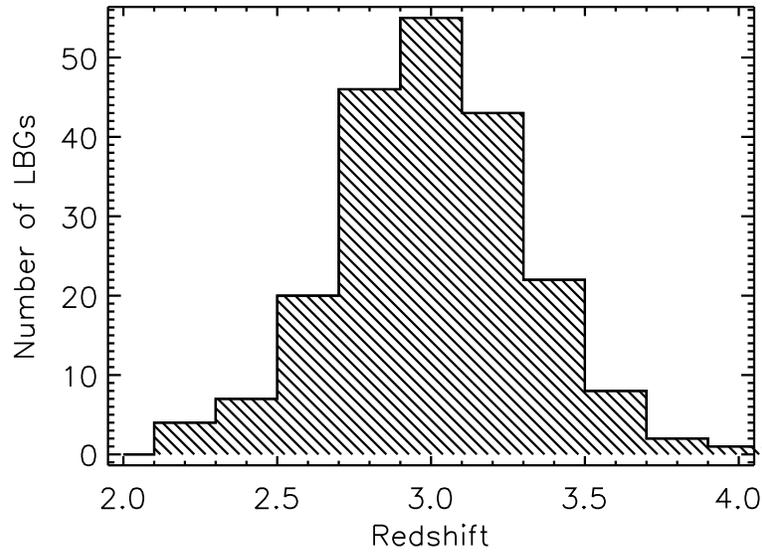}}
\caption
{Redshift distribution of the $z>2$ objects detected in this survey
  binned with $\Delta z = 0.2$.  The $u'$BVRI color selection
  technique results in a Gaussian distribution of spectroscopically
  confirmed objects centered at $z=3.02\pm 0.32$.}
\label{fig:zdistrib}
\end{center}
\end{figure}

Spectroscopic observations were obtained from 2000 November through
2004 February using LRIS on the Keck I telescope.  Spectra were
acquired either using the 300 line mm$^{-1}$ grating on the red arm,
the 300 line mm$^{-1}$ grism on the blue arm, or both.  We used
multi-object slitmasks with $1.''0-1''.5$ slitlets, chosen according
to the seeing conditions whenever possible, that provided a spectral
resolution of $9-12$\AA~FWHM.  We obtained spectroscopy of 529 LBG
color candidates that resulted in 339 redshift identifications,
largely a consequence of weather, instrument failures, and shortened
exposure times.  We conservatively identified 211 $z>2$ LBGs for use
in the analyses in this work.  Details regarding the efficiency of the
$u'$BVRI photometric selection and the subsequent redshift
identifications can be found in Paper I.  In short, low-resolution,
low signal-to-noise ratio spectra of this caliber result in redshift
identifications that vary in confidence.  The efficiency of identified
$z>2$ objects that met the photometric criteria was as low as 62\% for
stringent identifications and as high as 95\% when including the low
confidence identifications.  The redshift distribution of $z>2$ LBGs
used in the analysis is shown in Figure~\ref{fig:zdistrib}.

\subsubsection{LBG redshifts}\label{redshifts}

LBGs have prominent rest-frame UV spectral features from 912\AA~to
$\sim1700$\AA~\citep{pet00,aes03} that are redshifted to
$\sim3200$\AA$-7500$\AA~for objects at $z\sim3$.  These features are
identifiable in spectra with moderate to low signal-to-noise ratios.
All LBGs in this work were identified following the procedure
described in Paper I.  Redshifts were determined using Lyman $\alpha$
features, continuum profiles, and one to many stellar and interstellar
absorption and emission lines.  The observed discrepancy between the
Lyman $\alpha$ emission and interstellar absorption redshifts
witnessed to some extent in all LBG spectra to date is attributed to
the effects of galactic-scale stellar and supernovae-driven winds.
The uncertainty in systemic redshift caused by this discrepancy was
minimized by adopting the corrections outlined in A03 and presented
below.  These corrections were formulated from the results of
rest-frame optical nebular measurements of a sample of LBGs
\citep{pet01} with the justification that the gas responsible for LBG
nebular lines \textsc{[Oii]} $\lambda$3727, H$\beta$, and
\textsc{[Oiii]} $\lambda$$\lambda$4959, 5007 closely traces the
redshifts of the stellar populations.

For LBGs displaying Lyman $\alpha$ in emission, the correction to the
systemic redshift velocity determined solely from the observed Lyman
$\alpha$ emission feature is
\begin{equation}\label{eq:zcorr-LyaEW}
v_{Ly\alpha} \simeq + 670 - 8.9~W_\lambda~\mbox{\AA}^{-1}\mbox{km
s}^{-1},
\end{equation}
\noindent where $W_\lambda$ is the rest-frame equivalent width of the
Lyman $\alpha$ emission in \AA.  In cases where the redshift is
determined by the Lyman $\alpha$ feature and the equivalent width is
uncertain the corrections is
\begin{equation}\label{eq:zcorr-Lya}
v_{Ly\alpha} \simeq + 310~\mbox{km s}^{-1}.
\end{equation}
\noindent For LBGs displaying Lyman $\alpha$ and interstellar metal
absorption lines, the mean velocity is corrected by
\begin{equation}\label{eq:zcorr-absmed}
v_{abs} \simeq -0.114\Delta v + 230~\mbox{km s}^{-1}
\end{equation}
\noindent where $\Delta v$ is the velocity difference between the
Lyman $\alpha$ feature and the average of the interstellar absorption
lines.  Lastly, the velocity correction 
\begin{equation}\label{eq:zcorr-abs}
v_{abs} \simeq - 150~\mbox{km s}^{-1}
\end{equation}
is applied to the redshifts determined solely by the average of the
measured interstellar absorption lines.

These formulae are expected to diminish the redshift uncertainties
caused by galactic outflows to within an rms scatter of $\sim200$ km
s$^{-1}$ ($\Delta z\sim 0.002$ at $z\sim3$).  No correction was made
for the unknown peculiar velocities of the LBGs.  We used as much
information as possible from our set of 211 $z>2$ LBGs and nearly
always secured more than one absorption line for each LBG spectrum
(see Paper I).  We found a negligible effect on the resulting
correlation functions when using equation~\ref{eq:zcorr-Lya} and
\ref{eq:zcorr-abs} when compared to using
equations~\ref{eq:zcorr-LyaEW} and~\ref{eq:zcorr-absmed}.

\section{Spectra of Systems near DLAs}\label{assoc}   

\begin{figure}
\begin{center}                          
\scalebox{0.6}[0.6]{\includegraphics{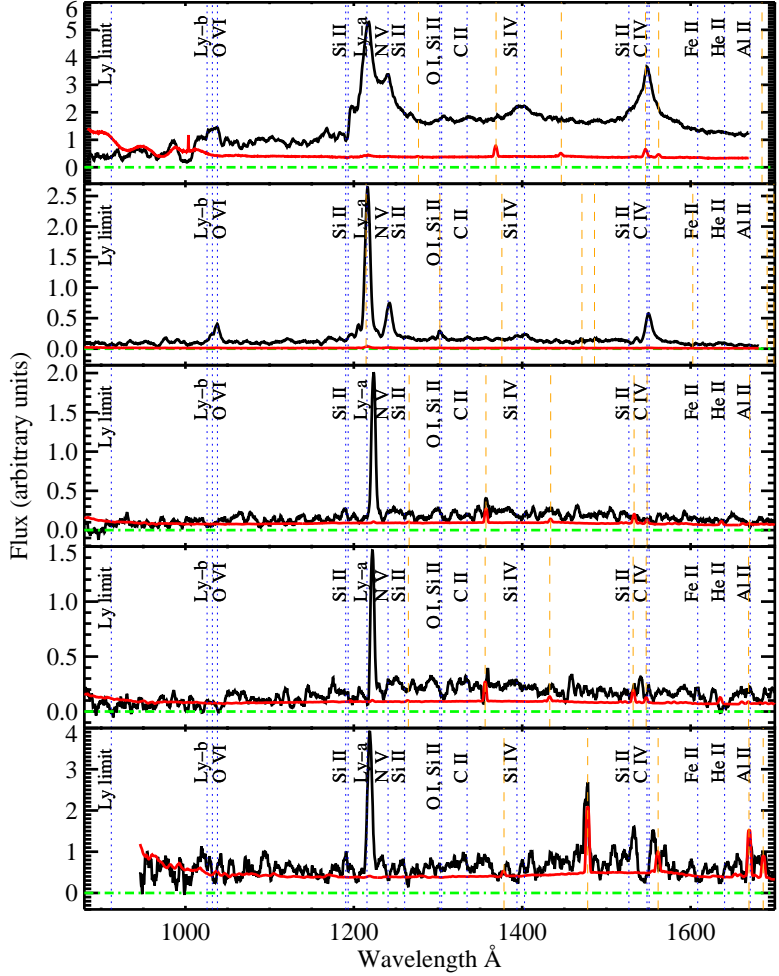}}
\caption{\small Figures~\ref{fig:assoc1-5rf} to~\ref{fig:assoc12-17rf}
  present individual spectra of the 13 Lyman break galaxies and 2 QSOs
  that reside within $\Delta z=0.0125$ of the 11 DLAs.  In addition,
  we include the spectra of two LBGs found within $\Delta z=0.015$ of
  the $z=2.936$ DLA in the PSS0808+5215 field.  Error arrays are
  overlaid (in red) and expected interstellar and stellar absorption
  lines are indicated using vertical dotted (blue) lines.  Bright
  night sky emission lines, that can be difficult to subtract
  completely in fainter spectra, are marked with vertical dashed
  (orange) lines. These spectra range from the lowest to highest
  signal-to-noise ratio of the complete sample.  They are smoothed by
  15 pixels for clarity over the wavelength range shown and to help
  highlight the break(s) in the continua, however the sharpness of the
  individual emission and absorption lines is affected.  Top to
  bottom: Objects 0336-0782 (QSO found near the DLA in the
  PKS0336--017 field), 0957-0859 (QSO found near the DLA in the
  PSS1057+4555 field), 1013-0210, 1013-0661, \& 0056-0993.}
\label{fig:assoc1-5rf}
\end{center}
\end{figure}

\begin{figure}
\begin{center}                          
\scalebox{0.6}[0.6]{\includegraphics{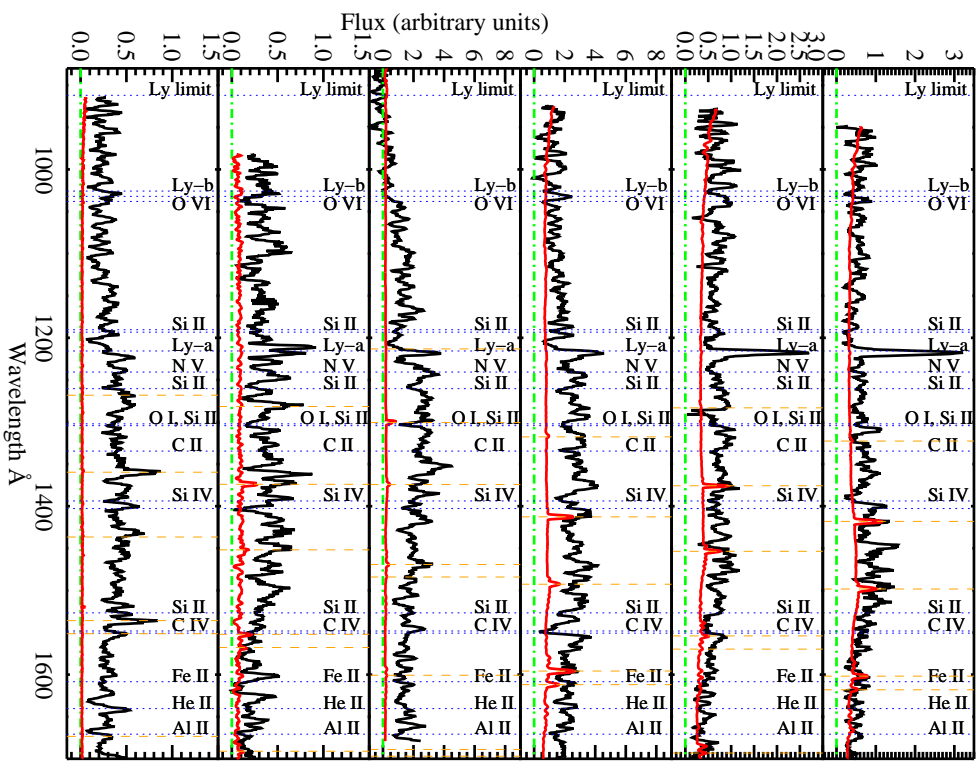}}
\caption{\small Top to bottom: Objects 0808-0944, 1057-1278,
  0808-0377, 0957-1284, 0808-0506, \& 0336-0758.}
\label{fig:assoc6-11rf}
\end{center}
\end{figure}

\begin{figure}
\begin{center}                         
\scalebox{0.6}[0.6]{\includegraphics{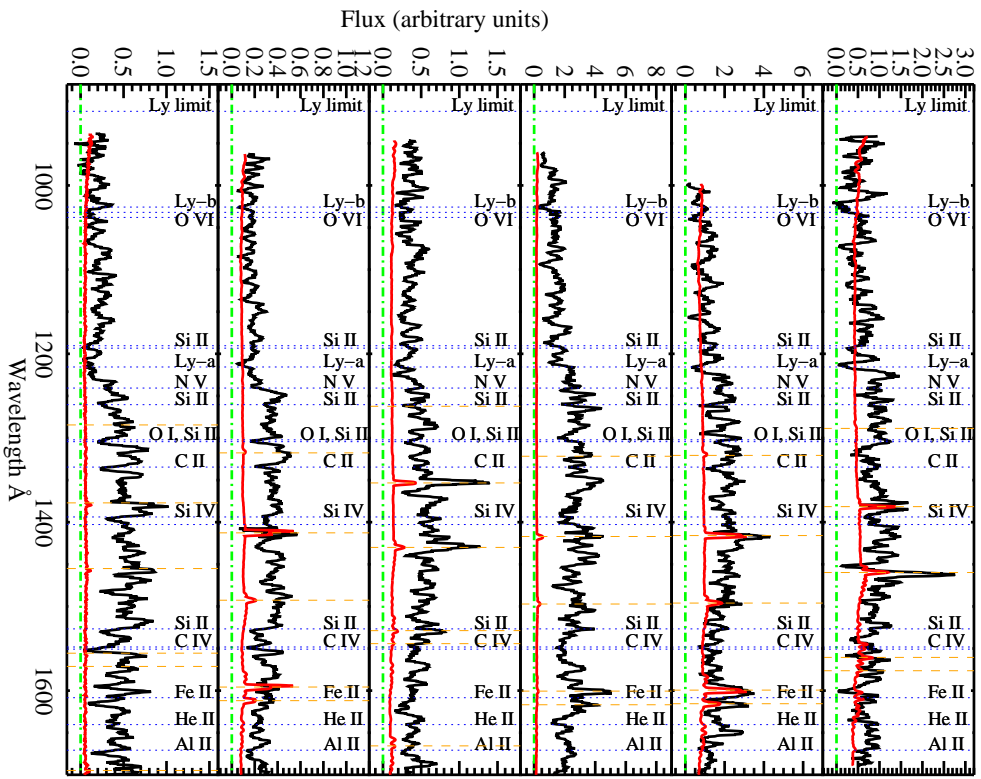}}
\caption{\small Top to bottom: Objects 1057-3646, 0808-1264,
0808-0269, 0808-0556, 0808-1801, \& 1057-0600.}
\label{fig:assoc12-17rf}
\end{center}
\end{figure}

Figures~\ref{fig:assoc1-5rf} through~\ref{fig:assoc12-17rf} present
the individual spectra of the 15 systems (13 LBGs and 2 QSOs) within
$\Delta z=0.0125$ of the 11 DLAs in this survey.  In addition, we have
included the spectra of two LBGs found within $\Delta z=0.015$ of the
$z=2.936$ DLA in the PSS0808+5215 field (see \S\ref{spike}).  The low
signal-to-noise spectra have been smoothed by 15 pixels.  This large
smoothing allows the coarse features of the continua to be seen more
readily on the wavelength scales presented here, but diminishes the
appearance of individual absorption features.  These range from the
highest to the lowest signal-to-noise ratio spectra in the complete
sample.  As presented in Paper I, these spectra are best referenced
to, and studied as, composite spectra of galaxies displaying similar
spectral profiles.

In nearly every spectra, a decrement in the continuum is visible
shortward of 1216\AA~caused by absorption from optically thick
intervening systems at lower redshift (the Lyman $\alpha$ forest).
LBGs are faint sky-dominated objects and bright night sky emission
lines can be difficult to subtract cleanly from the spectra.
Therefore, the positions of the sky emission lines are marked to
prevent the misidentification of residual sky flux as real LBG
features.  No order blocking filter was used in these observations
resulting in an underestimation of the flux longward of
$\sim6300$\AA~in the observation frame or longward of $\sim1500$\AA~in
the rest-frame.

\begin{figure}
\begin{center}                         
\scalebox{0.46}[0.43]{\rotatebox{90}{\includegraphics{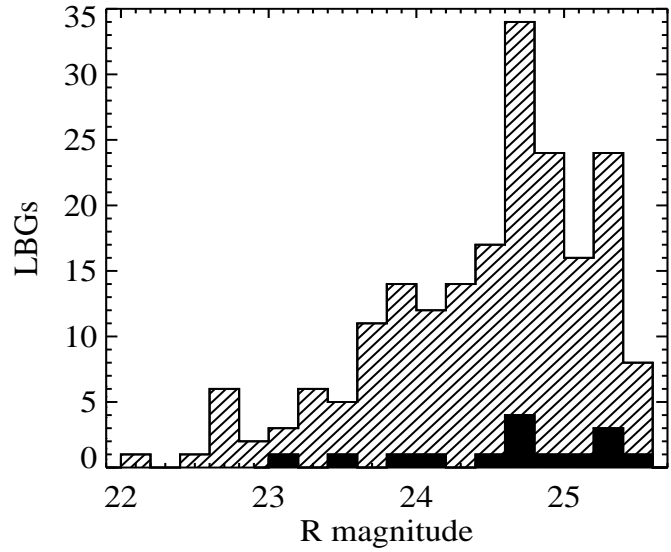}}}
\caption{Histogram of the R magnitude distribution of the full
  $2.6<z<3.4$ LBG sample (hatched) and the LBGs near DLAs (solid).  A
  two-sided Kolmogorov-Smirnov test revealed a high probability that
  the two cumulative distribution functions are significantly similar}
\label{Rmag}
\end{center}
\end{figure}

Overall, the spectra of the systems near DLAs appear to be those of
typical LBGs and we find a similar ratio of emission-identified LBGs
to absorption-identified LBGs as in the complete sample.  Excluding
QSOs, the 15 LBGs presented here exhibit a magnitude range of $23.1<$
R $<25.4$. The set of 205 LBGs with no apparent AGN activity has a
magnitude range of $22.1<$ R $<25.5$, where R $=25.5$ is the practical
spectroscopic magnitude limit of Keck using the LRIS instrument.  The
R magnitude distribution of these objects against the full set of LBGs
are shown in Figure~\ref{Rmag}.  A two-sided Kolmogorov-Smirnov test
resulted in a value of 0.6 and a high probability that the cumulative
distribution functions of both datasets are significantly similar.

\section{Faint AGN}\label{AGN}   
 
\begin{figure}
\begin{center}                          
\scalebox{0.6}[0.6]{\includegraphics{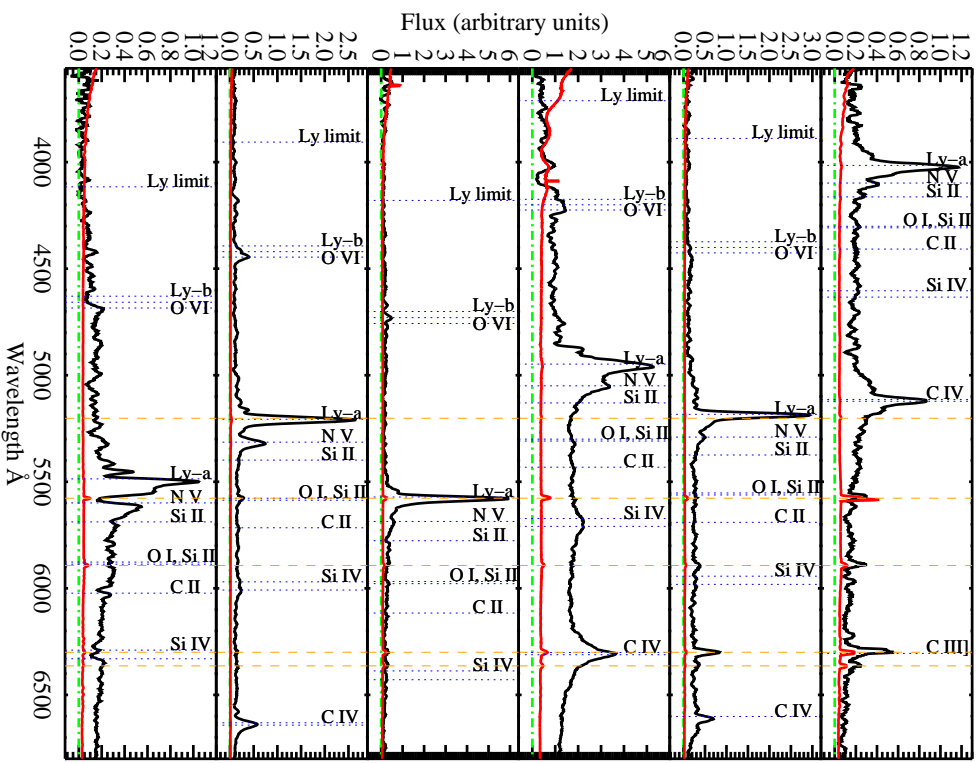}}
\caption{Spectra of the six $20.0<$ R $<25.5$ QSOs discovered in this
  survey with the error arrays are overlaid (in red).  Some identified
  emission and absorption lines are marked with vertical dotted (blue)
  lines and labeled.  Bright night sky emission lines are marked with
  vertical dashed (orange) lines.  Night sky lines can be difficult to
  subtract completely in the fainter spectra.  Top to bottom:
  0056-1085, 0336-0645, 0336-0782, 0808-0876, 0957-0859, \&
  1057-3225.}
\label{fig:faint-qsos}
\end{center}
\end{figure}

Figure~\ref{fig:faint-qsos} presents the smoothed spectra of six Lyman
break objects displaying AGN activity discovered in the 465 arcmin$^2$
of this survey.  These six $20.1<$ R $<24.4$ objects are separate and
distinct from the nine QSOs targeted in this survey and result in a
faint AGN number density of $\sim46\deg^{-2}$.  Three objects are
broad-line AGN with FWHM $>2000$ km s$^{-1}$ and display several broad
emission features that are typical of QSOs, and the remaining three
display emission lines with FWHM $<2000$ km s$^{-1}$ including Lyman
$\alpha$ and at least one other high ionization species indicative of
a hard spectrum.  The spectrum of object 0808-0876 did not extend to
\textsc{Civ} $\lambda\lambda 1548, 1551$, but detailed inspection does
show \textsc{Ovi} $\lambda\lambda1032, 1038$, \textsc{Nv} $\lambda
1240$, and possible \textsc{Siv} $\lambda\lambda1394, 1402$ emission.
As a result, this object can not be ruled out as a Lyman break galaxy
in the conventional sense of the term.

Interestingly, two of the six QSOs were discovered within $\Delta r_z=
0.0125$ ($<10h^{-1}$Mpc) of two DLAs.  Object 0957-0859, an R $=23.3$
narrow-line QSO at $z=3.283$, lies near the $z=3.279$ DLA in the
PSS0957+3308 field at an angular separation of 241 arcsec.  Object
0336-0782 at $z=3.074$ is a brighter R $=20.1$ broad-line QSO and lies
at an angular separation of 167 arcsec from the $z=3.062$ DLA in the
PKS0336--017 field. Since the appearance of two QSOs at small
separations from the two DLAs seemed unlikely, we tested this in the
following manner.  We assumed the detection of one QSO per survey DLA
field (we found six QSOs in nine fields) and ran a Monte Carlo
simulation of 10,000 realizations corrected by the photometric
selection function.  From this, we estimate a 3.8\% and 2.8\% chance
that a QSO would reside randomly within $\Delta r_z= 0.0125$ of the
$z=3.062$ and $z=3.279$ DLA, respectively. More importantly, we
estimate 1 chance in 940 that a QSO would reside within $\Delta r_z=
0.0125$ of {\it both} DLAs. This is significant to $\sim4\sigma$ and
may have important implications on the distribution of QSOs with
DLAs. The close proximity of the QSOs with the DLAs may provide
insight into the duty cycle of QSOs and the overall size and survival
of high column-density neutral gas reservoirs in environments with
sources of significant ionizing flux.  More research into this
relationship is necessary and is one of the goals of our ongoing
One-Degree Deep survey \citep{odd}.

In the following sections, we focus on the distribution of the LBG
population as a whole.  Throughout, we search for evidence of an
overdensity of LBGs near DLAs over random and compare this to the
overdensity of LBGs near other LBGs.  We describe the approach adopted
to estimate the three-dimensional LBG auto-correlation and DLA-LBG
cross-correlation functions and present several techniques to measure
and test these functions from the dataset.

\section{Correlation Functions: Methodology}\label{methodology} 

For a random distribution of LBGs, the joint probability of finding an
LBG occupying volume element $dV_1$ and another LBG occupying volume
element $dV_2$ at a separation $r$ is \citep{peebles80}
\begin {equation}\label{randomprob}
dP(r) = \bar n_{LBG}^2 ~ dV_1 dV_2
\end {equation}
where $\bar n_{LBG}$ is the mean density of LBGs averaged over the
realization.  In general, for any given distribution of LBGs, this 
expression becomes 
\begin {equation}\label{xiprob}
dP(r) = \bar n_{LBG}^2 ~ [1 + \xi_{LBG}(r)] dV_1 dV_2
\end {equation}
\noindent where $\xi_{LBG}(r)$ is the LBG auto-correlation function.
In this context, $\xi_{LBG}(r)$ quantifies the excess probability over
random.  

Similarly, for two populations (here DLAs and LBGs), the joint
probability of finding an object from the first population at a
distance $r$ from an object of the second population is
\begin {equation}
dP(r) = \bar n_{DLA} \bar n_{LBG}~[1 + \xi_{DLA-LBG}(r)]~dV_{DLA}
dV_{LBG}
\end {equation}
where $\bar n_{DLA}$ is the mean density of DLAs and
$\xi_{DLA-LBG}(r)$ is the cross-correlation function also quantifying
the excess probability over random.  From this, the conditional
probability of finding an LBG at a distance $r$ from a known DLA is
\begin {equation}
dP(r) = \bar n_{LBG}~[1 + \xi_{DLA-LBG}(r)]~dV_{LBG}.
\end {equation}

Based on studies of nearby galaxies it has been commonly assumed that
$\xi(r)$ follows a power law of the form
\begin {equation}\label{powerlaw}
\xi(r) = \left(\frac{r}{r_0}\right)^{-\gamma}.
\end {equation}
This has been a reasonable assumption given that the power spectrum is
well fit by a power law and that $\xi(r)$ is essentially the Fourier
transform of the power spectrum.  In this form, the parameters $r_0$
and $\gamma$ are all that are needed to describe the correlation
function. In practice, $\xi(r)$ is estimated by comparing the galaxy
separations found in the data to the galaxy separations in mock
catalogs of randomly distributed galaxies.  These random galaxy
catalogs mimic the angular and spatial configuration of the data
[e.g. \citet{davis83,hawkins03,a03}].  By carefully restricting the
random galaxy catalogs to the exact constraints of the real data,
complications caused by edge effects, bright objects, and the physical
constraints of the instruments are removed or well constrained.

\subsection{Spatial correlation estimator}\label{LS}

There have been several methods proposed and used to estimate $\xi(r)$
from galaxy catalogs.  We adopt the method of \citet{ls93} which is
well-suited for small galaxy samples and has the least bias present in
commonly used estimators \citep{k00}.  This technique involves
comparing the number of galaxy pairs in the data having separations
within a given spatial interval $r\pm\delta r$ to the number of galaxy
pairs in the random galaxy catalogs having separations within the same
spatial interval.  To reduce shot noise, the random galaxy catalogs
are made many times ($\sim100$ times) larger than the data sample and
normalized to the data.  The number of pairs is counted in each
spatial bin determined in logarithmic or linear space.  From the
normalized bin counts, the LBG auto-correlation function
$\xi_{LBG-LBG}~(r)$ is estimated as
\begin {equation}\label{LBGest}
\xi_{LBG-LBG}~(r) = \frac{D_{LBG}D_{LBG} - 2\cdot D_{LBG}R_{LBG} +
  R_{LBG}R_{LBG}}{R_{LBG}R_{LBG}}
\end {equation}
and the DLA-LBG cross-correlation function $\xi_{DLA-LBG}(r)$ is
estimated as 
\begin {equation}\label{DLAest}
\xi_{DLA-LBG}~(r) = \frac{D_{DLA}D_{LBG} - D_{DLA}R_{LBG} -
  R_{DLA}D_{LBG} + R_{DLA}R_{LBG}}{R_{DLA}R_{LBG}}
\end {equation}
\noindent where the separations between galaxies in the data
constitute the $DD$ catalogs, separations between random galaxies make
up the $RR$ catalogs, and separations between data and random galaxies
make up the $DR$ and $RD$ cross-reference catalogs.
Equations~\ref{LBGest} \&~\ref{DLAest} are identically used to
estimate the projected angular correlation functions $\omega_p$ in
\S~\ref{adel_scheme}.

\section{Correlation Functions: Results by Technique}\label{results}   

We present several approaches to measure and test for an overdensity
of LBGs near LBGs and LBGs near DLAs over random.  We first describe
the correlation functions as determined by a conventional binning
technique.  We find a dependence of the correlation function on bin
parameters and circumvent this shortcoming, which can be pronounced
for small datasets, by performing a maximum likelihood analysis and
comparing the results.  The maximum likelihood method makes the most
of the dataset and is a direct and essentially bin-independent way to
determine the clustering behavior.  Lastly, we test the effects that
the physical constraints of the slitmasks and the presence of an
individual overdense field have on the correlation function.  The
redshift separations in all analyses were determined in a consistent
manner.

\subsection{Conventional binning}\label{adel_scheme}

We followed the modification to conventional radial bins suggested by
A03 in an effort to diminish the effects that the LBG redshift
uncertainties caused by galactic-scale winds have on the clustering
amplitude.  In doing so, we also provide a means for direct comparison
by methodology of our measure of the LBG auto-correlation and the
DLA-LBG cross-correlation functions to the LBG auto-correlation
function of A03.  In this treatment, the number of pairs is counted
that reside in concentric ``cylindrical'' bins with dimensions
$r_{\theta}\pm \delta r_{\theta}$ and $r_z \pm \delta r_z$. Limits are
placed on $r_z$ such that it is the greater of $7 r_\theta$ and 1000
km sec$^{-1} (1+ z)/H(z)$ and chosen to be several times larger than
the redshift uncertainties.  Here we interpreted $(1+z)$ as $(1+\bar
z)$ where $\bar z$ is the average redshift of the two objects.  The
length in redshift of each bin is fixed ($\sim9h^{-1}$Mpc at $z\sim3$)
for small $r_\theta$ and grows as $7r_\theta$ when $r_\theta$ becomes
large.  The lower limit is placed to avoid missing correlated pairs
and the upper limit reaches down the correlation function to include
$>80\%$ of the correlated pairs for $\gamma\gtrsim1.6$.

We measured out to the maximum angular separation of $\theta=300''$
and used logarithmic $r_{\theta} \pm dr_{\theta}$ bins to remain
consistent with the parameters chosen by A03.  Recalling that the
fields in this survey are $\sim6'\times7.5'$, the number of pairs with
separations at larger radii diminishes rapidly.  Assuming the
canonical power law form of the correlation function
(equation~\ref{powerlaw}), the expected excess number of pairs using
this approach is
\begin{equation}\label{omegap}
\omega_p(r_\theta ,<r_z)\equiv \frac{\langle n \rangle}{\bar n} - 1 =
\frac{r_0^\gamma r_\theta^{1-\gamma}}{2r_z} B \left(
\frac{1}{2},\frac{\gamma - 1}{2} \right) I_x
\left(\frac{1}{2},\frac{\gamma - 1}{2} \right) 
\end{equation}
where $\langle n \rangle$ is the expected number of objects, $\bar n$
is the number of objects in a random sightline, $B$ and $I_x$ are the
beta and incomplete beta functions with $x\equiv r_z^2 (r_z^2 +
r_\theta^2)^{-1}$ (Press et al. 1992, \S6.4).  The best fit values of
$r_0$ and $\gamma$ of the correlation function result from fitting
equation~\ref{omegap} to the observed number of pairs measured by the
above binning scheme.  The fundamental errors in $\omega_p$ are
dependent on our choice of estimator.  Using the estimators in
equations~\ref{LBGest} \&~\ref{DLAest}, the errors $\delta \omega_p
(\theta)$ are described by \citep{roche99,foucaud03}
\begin{equation}\label{LSerror}
\delta \omega_p (\theta) =  \sqrt{\frac{1 + \omega_p (\theta)}{DD}}
\end{equation}
where $DD$ represents either the $D_{LBG}D_{LBG}$ or $D_{DLA}D_{LBG}$
pair catalogs.

The uncertainties of the functional fits were estimated by running a
Monte Carlo simulation of the measured correlation function.  As
described in Appendix C, A03, this approach is to create a large
number (we performed 1000) of realizations of $\omega_p$ by adding a
Gaussian deviate to the fundamental error and minimizing the $\chi^2$
fit.  The range of 68\% of the best fit parameter values is what we
report as the best fit value $1 \sigma$ uncertainty when using this
method.  These uncertainties may be underestimated by a factor of
$1-2$, as argued in \citet{a05} and \citet{kurt05}.

\subsubsection{LBG auto-correlation}\label{lbg_ac}

\begin{figure}
\begin{center}                          
\scalebox{0.32}[0.28]{\rotatebox{90}{\includegraphics{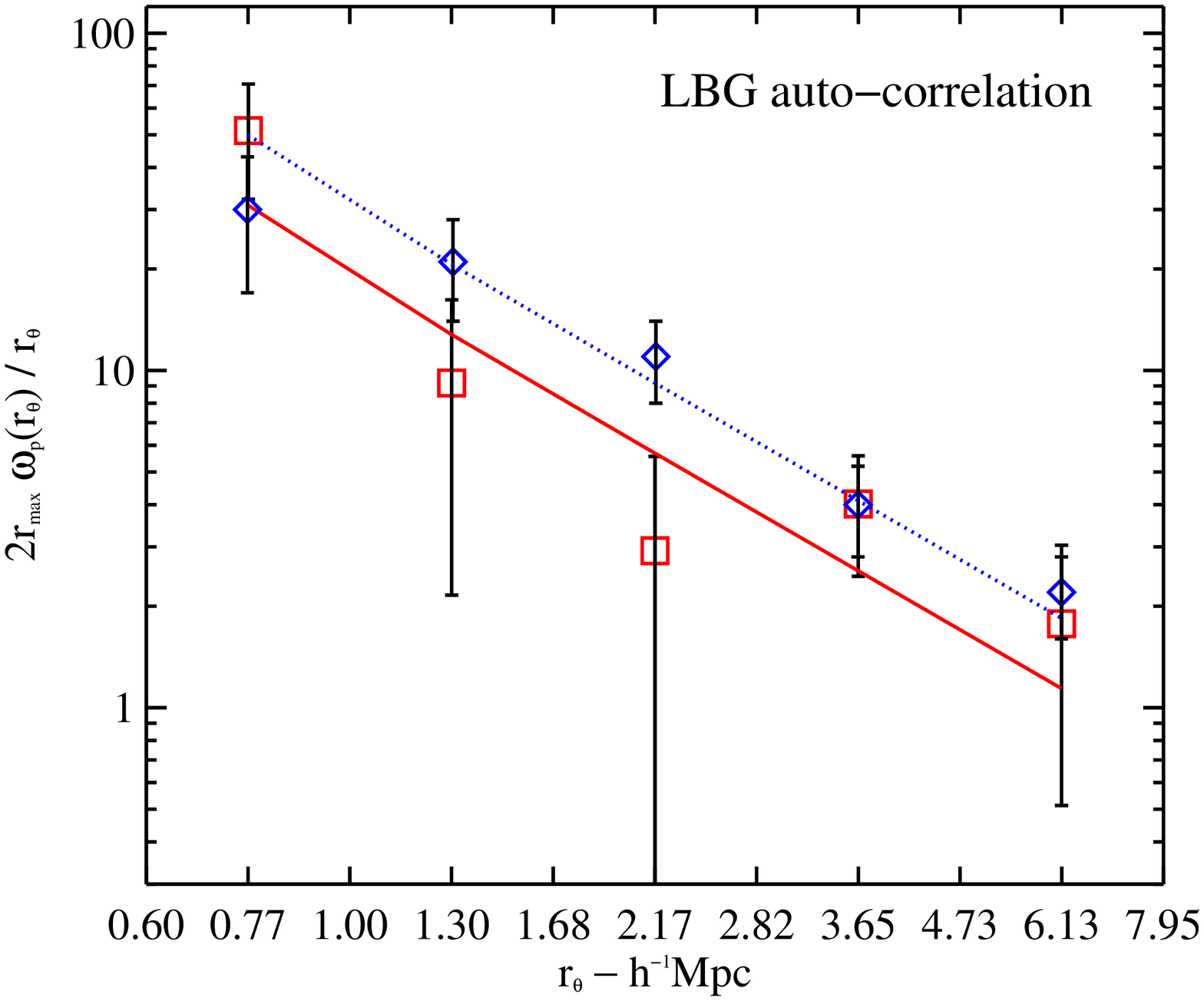}}}
\end{center}
\begin{center}                          
\scalebox{0.32}[0.28]{\rotatebox{90}{\includegraphics{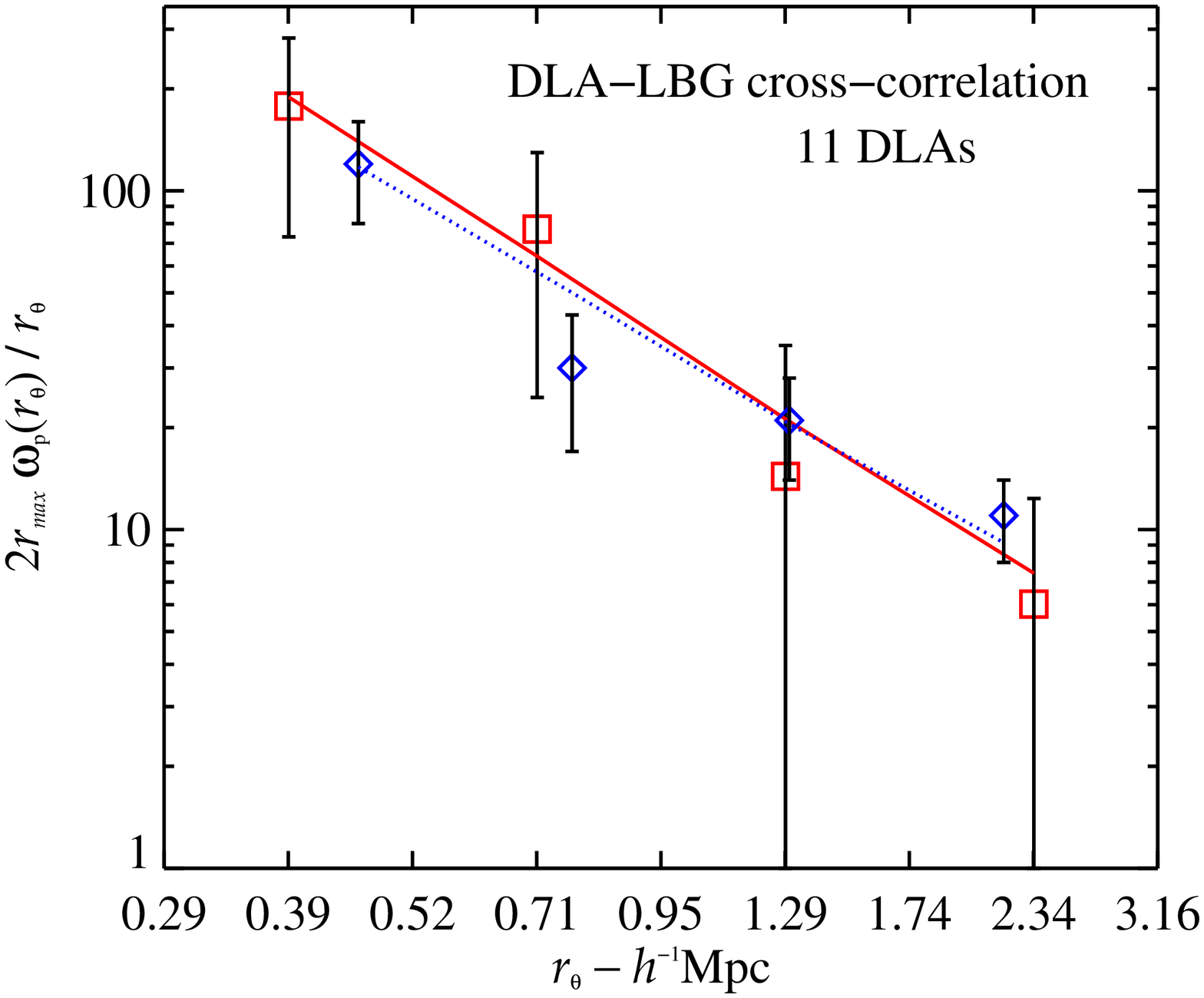}}}
\end{center}
\begin{center}                          
\scalebox{0.32}[0.28]{\rotatebox{90}{\includegraphics{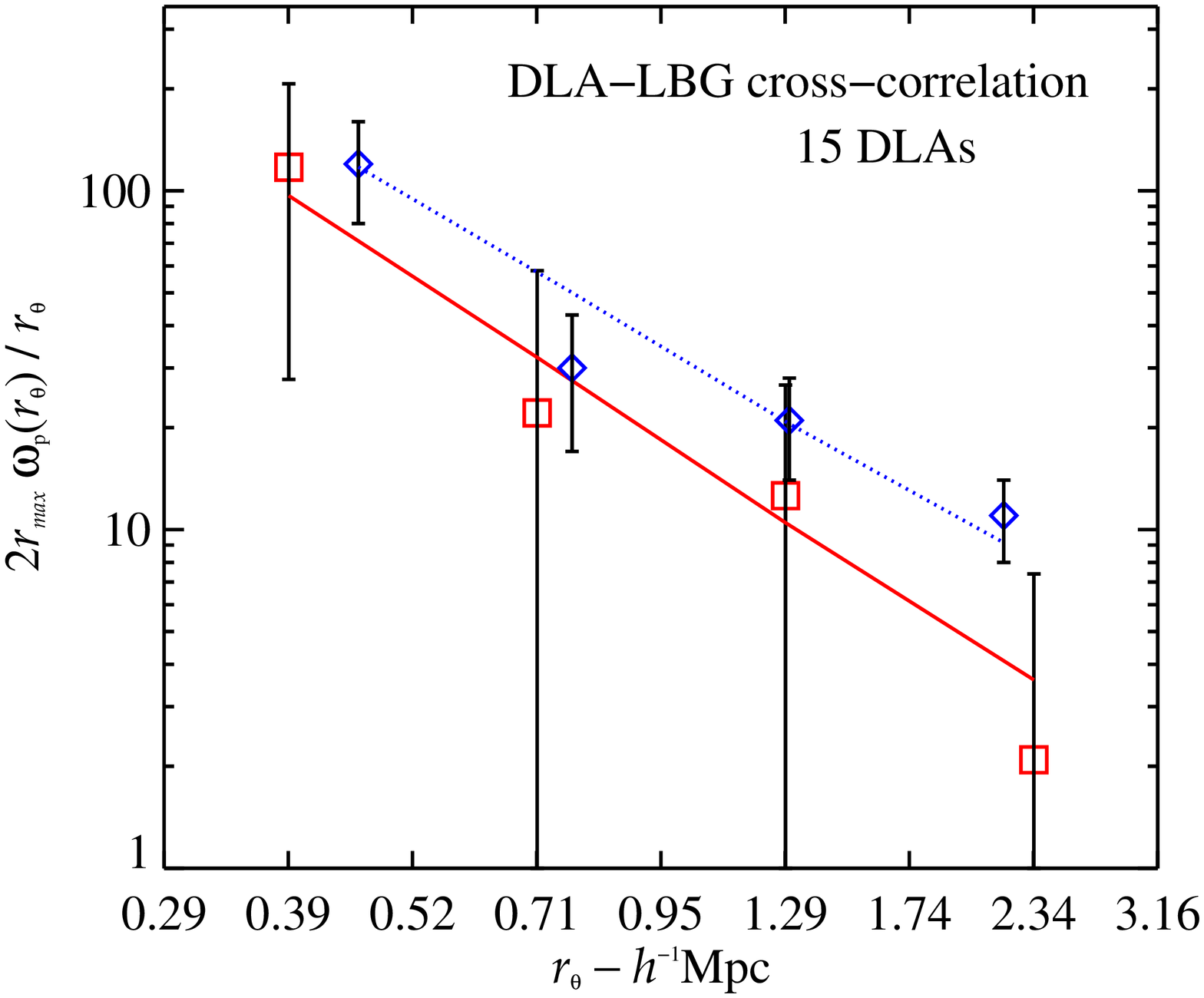}}}
\caption
{\small Plot of the LBG auto-correlation and DLA-LBG cross-correlation
  functions using the binning technique described in the text.  The
  correlation functions are shown as (red) squares and best fit as the
  solid (red) lines. We measure the best fit values and $1 \sigma$
  errors of $r_0=2.65\pm 0.48, \gamma=1.55\pm 0.40$ for the LBG
  auto-correlation (upper panel), $r_0=3.32\pm 1.25, \gamma=1.74\pm
  0.36$ for the DLA-LBG cross-correlation of the 11 DLAs in this
  survey (center panel), and $r_0=2.70\pm 1.16, \gamma=1.73\pm0.39$
  for the DLA-LBG cross-correlation of the combined set of 15 DLAs
  (lower panel) that includes four DLAs from the survey of
  \citet{s03}.  The errors shown are those determined by
  equation~\ref{LSerror}.  Overlaid in (blue) diamonds are the data
  for the LBG auto-correlation from \citet{a03} and best fit to that
  data, marked by the dotted (blue) lines, for comparison.}
\label{fig:LL}
\end{center}
\end{figure}

A03 reported the values and $1 \sigma$ uncertainties of $r_0=3.96\pm
0.29$ and $\gamma=1.55\pm 0.15$ for the LBG auto-correlation at
$z\sim3$.  As stated above, we fit our data in an identical manner as
A03 including the Monte Carlo error analysis.  The best fit values and
$1 \sigma$ uncertainties for our dataset are $r_0=2.65\pm 0.48$ and
$\gamma=1.55\pm 0.40$ and is shown in Figure~\ref{fig:LL}, with the
published results of A03 overlaid for comparison. The bin errors shown
in the figure (and all subsequent figures by this method) are the
error estimates on $\omega_p$ using equation~\ref{LSerror}.  We find a
possibly weaker clustering amplitude for $z\sim3$ LBGs in our sample
as compared to A03, yet consider the possible error underestimation on
the functional fit to the data as mentioned above. In addition,
subtleties involved in random catalog generation, sample variance and
sample size, estimation of the $U_nG\cal R$I versus $u'$BVRI
photometric selection function profiles, LBG redshift assignments, and
our inability to accurately measure the correlation function at
separations smaller than $\sim0.5 h^{-1}$Mpc may contribute as well.
In the remaining plots, all comparisons of the best fit values for the
LBG auto-correlation and the DLA-LBG cross-correlation must consider
these possible differences.

\subsubsection{DLA-LBG cross-correlation}\label{DL_cc}

We performed the above binning technique on the cross-catalogs of DLA
and LBG separations to determine the first spectroscopic measure of
the DLA-LBG cross-correlation function.  The best fit to the
cross-correlation data resulted in values and $1 \sigma$ uncertainties
of $r_0=3.32\pm 1.25$ and $\gamma=1.74\pm 0.36$ \citep{c06} and is
shown in Figure~\ref{fig:LL}. Upon inspection, it is immediately
apparent that the DLA-LBG cross-correlation function has a similar
slope and correlation length as the LBG auto-correlation function. The
angular range of the plot ($\sim0.4-3 h^{-1}$Mpc) reflects the limits
on the correlation measurement caused by placing the DLAs in the
center of our images.  Although the uncertainty in either
cross-correlation parameter is large, the measured central values
indicate an overdensity of LBGs near DLAs to $1-2 \sigma$.

\subsubsection{Inclusion of previous work}

The data from the four DLAs in the survey of \citet{s03}, hereafter
S03, and the 11 DLAs from this work constitute the largest available
spectroscopic sample of DLAs and LBGs to measure the spatial DLA-LBG
cross-correlation function.  The similarity in techniques and
instruments used in both surveys allowed a direct combination of the
data once the few differences were addressed and corrected to the best
of our abilities.  We used the available online dataset\footnote{Files
obtained from:
http://vizier.cfa.harvard.edu/viz-bin/VizieR?-source=J/ApJ/592/728/}
of S03 and note that our knowledge of some aspects of those
observations were limited.  In lieu of Lyman $\alpha$ equivalent width
information for each LBG in their sample, we were restricted to using
equations~\ref{eq:zcorr-Lya} and~\ref{eq:zcorr-abs} to determine the
systemic redshifts of 880 LBGs in 17 fields [and later a sub-sample of
700 in 15 fields from \citet{a04}].  The area of each S03 field was
estimated by the extent of the angular positions of their
spectroscopic data (this assumes a position angle identical to, or
having a right angle to, PA=0 for each slitmask).  Random catalogs
were generated in the same manner as described above using these field
sizes and the observed density of their sample.

Once we were satisfied with our duplication of the LBG
auto-correlation of A03, we measured the best fit values and $1
\sigma$ uncertainties of $r_0=2.20\pm 0.96$, $\gamma=1.77\pm0.40$ for
the DLA-LBG cross-correlation for the full set of 15 DLAs.  The
results are presented in Figure~\ref{fig:LL}.  It can be seen that
evidence for an overdensity of LBGs near DLAs survives and,
acknowledging the above caveats and our efforts to correct the
differences between the two surveys, the 15 DLAs may provide a better
sample to determine the cross-correlation parameters of DLAs at
$z\sim3$.

\subsection{Maximum likelihood}\label{ml}

\begin{figure}
\begin{center}                         
\scalebox{0.4}[0.4]{\rotatebox{90}{\includegraphics{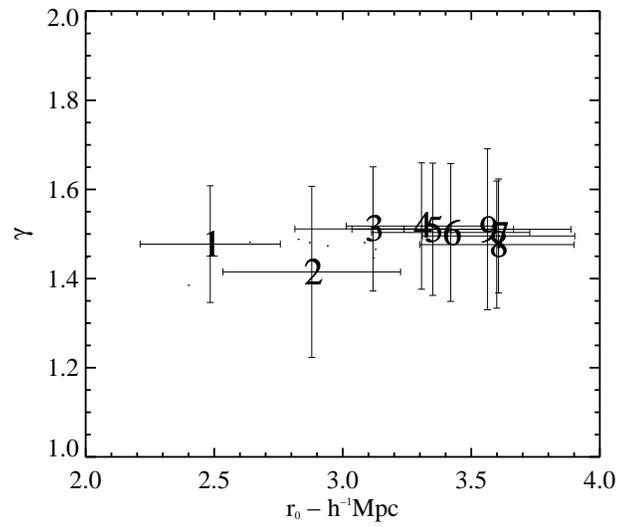}}}
\caption
{Scatter plot of the LBG auto-correlation from our analysis of the
  dataset of \citet{s03}.  We measured the parameters while varying
  the logarithmic bin size from $1.125$ to $0.225$ (labeled $1-9$)
  over a fixed interval of $r_\theta\sim0.04-8 h^{-1}$Mpc.  The effect
  of bin size on the correlation values is readily apparent.  Although
  the values appear to converge, smaller samples only allow a few
  larger bins and may not converge. }
\label{fig:binscatter}
\end{center}
\end{figure}

Arbitrary binning of the data into coarse bins introduces
uncertainties because the value of $\xi(r)$ can depend on the bin
size, interval, and bin center.  Dependence on bin size is illustrated
in Figure~\ref{fig:binscatter}.  In that plot, we varied the bin size
from logarithmic intervals of $0.225$ to $1.125$ over a fixed
$r_\theta\sim0.04-8h^{-1}$Mpc and found that $r_0$ varied from
$\sim2.0-3.1 h^{-1}$ Mpc and $\gamma$ varied from $\sim1.35-1.48$ in
our analysis of the S03 sample.  To remedy this problem, we estimated
the value of $r_0$ and $\gamma$ in the most direct way possible.  We
maximized the likelihood that a power law of the form $\xi(r) =
(r/r_0)^{-\gamma}$ would produce the observed pair separations
\citep{croft97,mullis04}.

Poisson probabilities are valid in the limit where the number of bins
is large and the probability per bin is small.  Constructing bin
separations on such a fine level as to include either one or zero LBGs
per bin allows us to form the likelihood function.  The probabilities
associated with the bins are assumed independent of each other in the
sparse sampling limit.  The probability $P$ of finding $\nu_i$
observed pairs where $\mu_i$ pairs are expected is the Poisson
probability
\begin{equation}
P_{\mu_i}(\nu_i) = \frac{e^{-\mu_i} \mu_i^{\nu_i}}{\nu_i !}.
\end{equation}

The likelihood function $\cal L$ is the product of the probability of
having exactly one pair in every interval where one pair exists in the
data and exactly zero in all others and is defined in terms of the
joint probabilities
\begin{equation}
{\cal L} = \prod_i^N \frac{e^{-\mu_i} \mu_i^{\nu_i}}{\nu_i !} 
\prod_{j\ne i}^N  \frac{e^{-\mu_j} \mu_j^{\nu_j}}{\nu_j !}
\end{equation}

\noindent where $\mu_i$ is the expected number of pairs in the
interval $dr$, $\nu_i$ is the observed number of pairs for that same 
interval, and the index $j$ runs over the elements where there are no
pairs.  This can also be expressed as 
\begin{equation}
\ln {\cal L} = \sum_i^N (-\mu_i + \nu_i \ln \mu_i - \ln \nu_i !)
+ \sum_{j\ne i}^N (-\mu_j + \nu_j \ln \mu_j - \ln \nu_j !).
\end{equation}

\noindent The expected number of objects $\mu_i$ for a given radial
separation is obtained by solving equations~\ref{LBGest}
\&~\ref{DLAest} for $D_{LBG}D_{LBG}$ and $D_{DLA}D_{LBG}$,
respectively.  As stated above, we used the assumption that
$\xi(r)=(\frac{r}{r_0})^{-\gamma}$ (equation~\ref{powerlaw}) and
varied the values of both $r_0$ and $\gamma$ to determine the values
of maximum likelihood.  The maximum likelihood was determined by
minimizing the conventional expression
\begin{equation}
S = -2 \ln {\cal L}
\end{equation}
\noindent and using $\Delta S = S(r_{best},\gamma_{best}) - S(r_0,
\gamma)$ to determine $\chi^2$ confidence levels, observing that the
values of $S$ had $\chi^2$ distributions.

The maximum likelihood technique is a powerful tool in measuring the
likelihood of a given functional fit to the data but has at least one
shortcoming in the form presented here. As mentioned above, the
Poisson approximation is valid in the regime large interval number
(very small separation radius) and low probability.  But, even large
random catalogs will occasionally find zero pairs in the very small
intervals where this approximation is most accurate.  In these cases,
the likelihood may be less accurate or undefined. Therefore, we
imposed the following two conditions \citep{mullis05}: (1) the number
of separations in the data-random cross-catalogs $DR$ (and $RD$) must
be greater than zero for each bin which indirectly imposes the same
condition on the random-random catalogs $RR$ since they are larger,
and (2) imposing the constraint that $[\xi(r)-1] > 2DR/RR$ (or
$[\xi(r)-1] > (DR+RD)/RR$ for the cross-correlation). For the few
values of $r$ where this criteria was not met the expected value of
$DD$ was interpolated across the finely spaced intervals.
Interpolating the few instances where this occurred had little impact
on the final result because there were far fewer of these intervals
when compared to the total used for analysis.

\subsubsection{LBG auto-correlation}

\begin{figure}
\begin{center}                          
\scalebox{0.3}[0.28]{\rotatebox{90}{\includegraphics{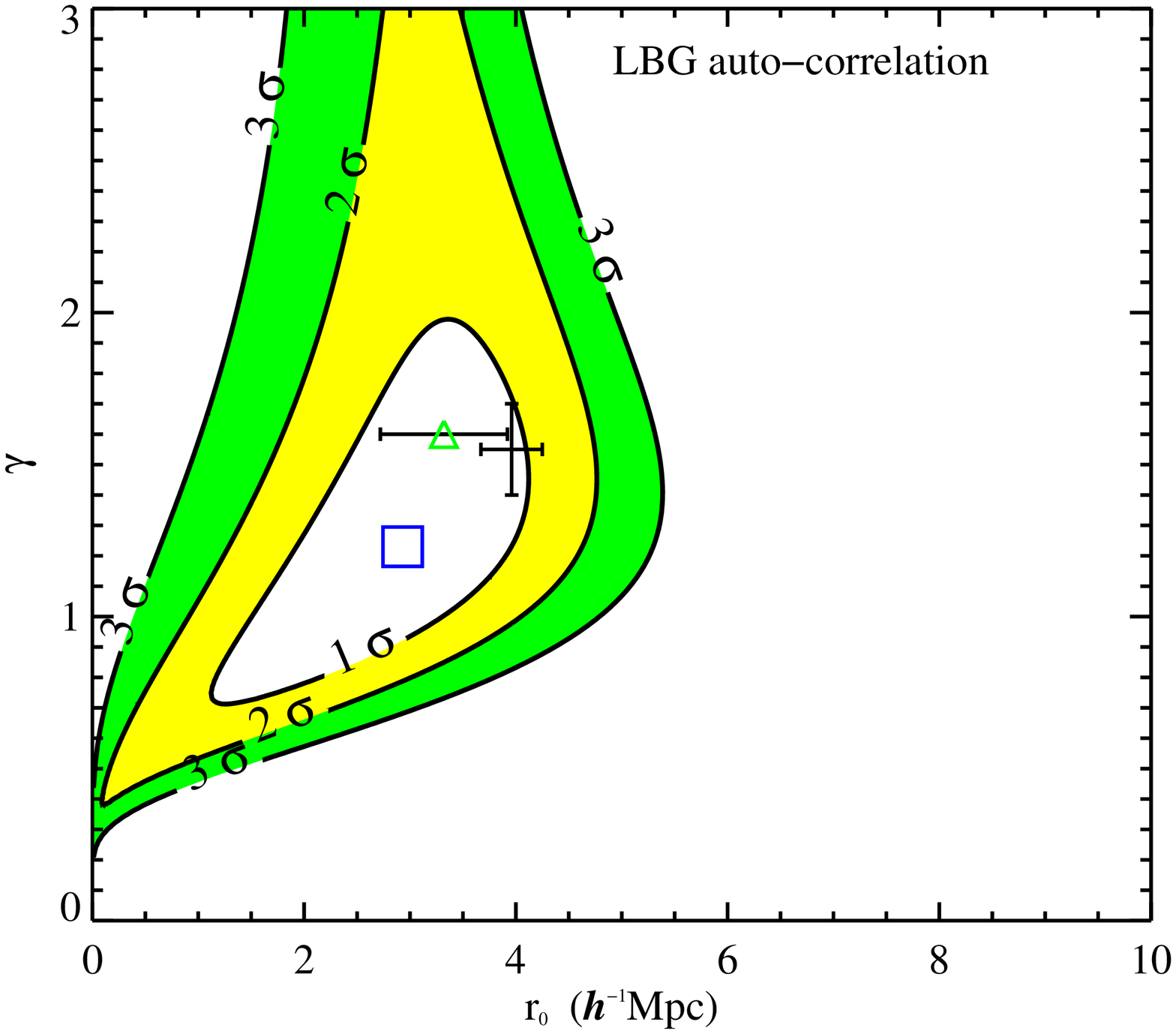}}}
\end{center}
\begin{center}                          
\scalebox{0.3}[0.28]{\rotatebox{90}{\includegraphics{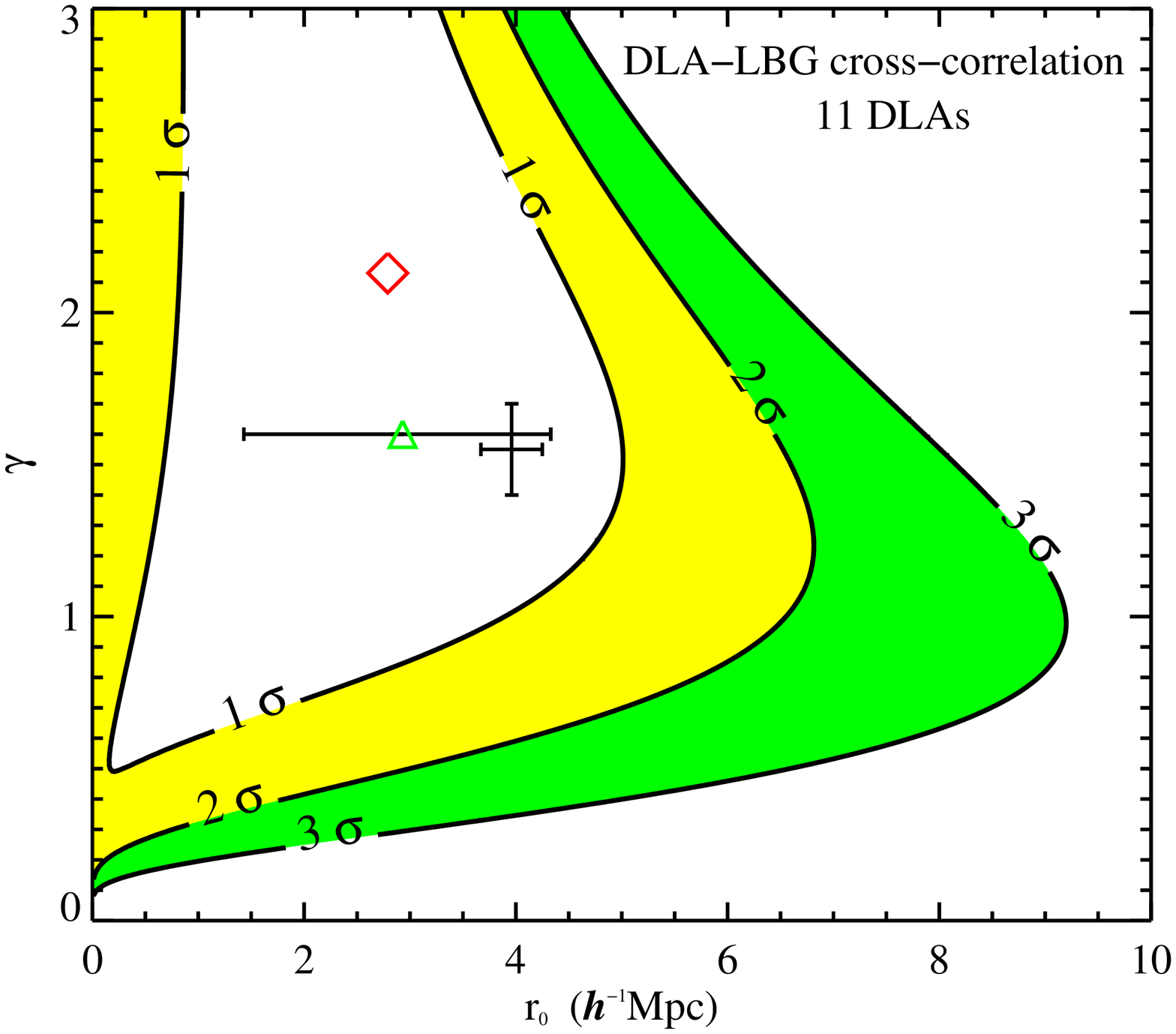}}}
\end{center}
\begin{center}                          
\scalebox{0.28}[0.27]{\rotatebox{90}{\includegraphics{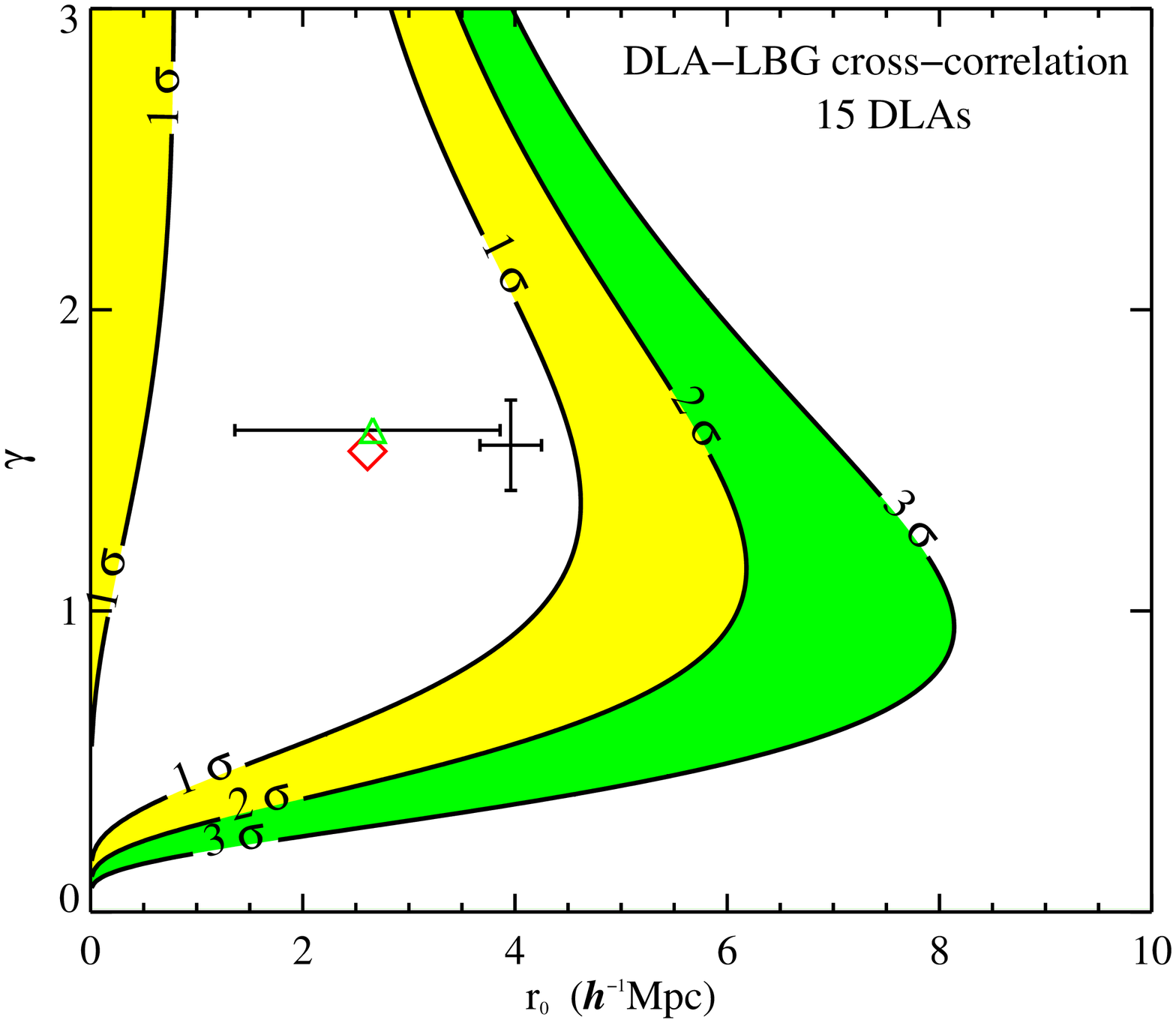}}}
\caption
{\small Correlation function probability contours for two free
  parameters as measured by the maximum likelihood method.  Top panel:
  The (blue) square at $r_0=2.91$, $\gamma=1.21$ marks the maximum
  likelihood values for the LBG auto-correlation for our
  survey. Middle panel: The (red) diamond marks the maximum likelihood
  values of $r_0=2.81$ and $\gamma=2.11$ for the DLA-LBG
  cross-correlation. Bottom panel: The (red) diamond marks the maximum
  likelihood values $r_0=2.66$, $\gamma=1.59$ for the DLA-LBG
  cross-correlation for the combined sample of 11 DLAS from this
  survey and the four DLAs from the survey of \citet{s03}.  The
  (green) triangles on each plot and $1 \sigma$ error bars are the
  best fit values (one free parameter) of $r_0$ for a fixed value of
  $\gamma=1.6$.  The error crosses indicate the values and $1 \sigma$
  errors for the LBG auto-correlation as determined by \citet{a03}.
  The 1, 2, and 3 $\sigma$ confidence regions are labeled.}
\label{fig:ml}
\end{center}
\end{figure}

The maximum likelihood values for the LBG auto-correlation and $1
\sigma$ confidence levels were found to be $r_0 = 2.91^{+1.0}_{-1.0}$
and $\gamma = 1.21`^{+0.6}_{-0.3}$.  The probability contours are
shown in the top panel Figure~\ref{fig:ml} where, in addition, we
overlay both the maximum likelihood value and $1 \sigma$ confidence
level of $r_0$ for a fixed $\gamma=1.6$ and the best fit values and $1
\sigma$ errors of the LBG auto-correlation function of A03 for
comparison.  The values of $r_0$ and $\gamma$ determined by this
method are consistent to within their errors with those found using
conventional binning, Moreover, the maximum likelihood technique
yields the same results regardless of the number intervals tested.

\subsubsection{DLA-LBG cross-correlation} 

Since the maximum likelihood method is well-suited for small samples,
we readily applied it toward the DLA-LBG cross-correlation
measurement.  The analysis found maximum likelihood values and
$1\sigma$ confidence levels of $r_0 = 2.81^{+1.4}_{-2.0}$ and $\gamma
= 2.11^{+1.3}_{-1.4}$ for the set of 11 DLAs and $r_0 =
2.66^{+1.9}_{-2.1}$ and $\gamma = 1.59^{+1.6}_{-0.9}$ for the full set
of 15 DLAs, with the probability contours shown in Figure~\ref{fig:ml}
(center and bottom panel, respectively).  We found that $65-90$\% of
the maximum likelihood values indicate a non-zero $r_0$ depending on
the value of $\gamma$.  Although the uncertainties are large, the best
fit values using the maximum likelihood technique also suggest an
overdensity of LBGs near DLAs with $1-2 \sigma$ confidence similar to
the results using conventional binning.

\subsection{Tests}\label{tests}

This survey uses the distribution of LBGs determined from multi-object
spectroscopic data to measure the correlation functions.  Here we test
the contributions to the correlation functions by the physical
constraints of the multi-object slitmasks and test the strength of the
clustering signal in the absence of the DLA having the largest
overdensity of LBGs.

\subsubsection{Physical constraints of the observations}

A false enhancement of the clustering signal can occur when the finite
number of multi-object slitmasks do not cover the full area of the
imaged fields.  This effect can be problematic in every survey and is
nearly removed here by the fact that seven out of the nine fields were
imaged with the relatively small field-of-view LRIS camera and have
spectroscopic coverage over their entire area.  The remaining two
fields, imaged by the larger field-of-view COSMIC, have $\sim70\%$
areal spectroscopic coverage.  However, any augmentation to the
correlation functions from these two fields was virtually eliminated
by confining our random catalogs to the precise areas sampled by the
slitmasks and by the fact that these two fields have few spectra and
make a small contribution to the final results.

Perhaps a more prominent effect is the dilution to the clustering
signal from the fact that only a finite number of objects are allowed
on each slitmask.  All but one of the LBG candidates that lie in
conflict in the dispersion direction are compromised.  Similarly, LBGs
that cluster tightly in angular space require many slitmasks for
proper spectroscopic coverage which is not usually feasible.  In order
to minimize this, we observed two to three overlapping slitmasks in
most fields.  Even so, there remain a few tightly clustered LBG
candidates as well as LBG candidates that were in conflict in the
dispersion direction that have no spectral coverage to date.  To
measure the extent in which this physical constraint affects the
clustering signal, we compared the results of the correlation
functions using random catalogs having galaxies with the exact angular
positions of the data to those using random catalogs having galaxies
with random angular positions.

The correlation measurements presented in \S\ref{adel_scheme}
\&~\S\ref{ml} used random galaxy catalogs having the exact angular
positions of the data.  We re-measured the correlation functions using
these techniques but allowed the galaxies in the random catalogs to
have random angular positions.  Doing so, we found the best fit
parameters and $1 \sigma$ uncertainties for the LBG auto-correlation
to be $r_0=2.31\pm 0.55$, $\gamma=1.47\pm 0.40$ for the conventional
binning method and $r_0=2.08^{+1.0}_{-1.1}$,
$\gamma=1.49^{+1.1}_{-0.5}$ for the maximum likelihood method.
Duplicating this for the DLA-LBG cross-correlation of the 11 DLAs in
our survey [and the combined set of 15 DLAs], we found the best fit
parameters and $1 \sigma$ uncertainties to be $r_0=3.21\pm 0.95$,
$\gamma=2.03\pm 0.22$ [$r_0=2.52\pm 0.92$, $\gamma=1.71\pm 0.46$]
using conventional binning and $r_0=3.20^{+2.2}_{-2.9}$,
$\gamma=1.62^{+1.4}_{-1.0}$ [$r_0=2.44^{+1.3}_{-1.9}$,
$\gamma=2.41^{-1.6}_{-1.7}$] using the maximum likelihood technique.
There was no apparent trend in either parameter which suggests that
the physical constraints of the slitmasks had a weak effect on our
survey as a whole.  This was suspected since, in most cases, we
obtained overlapping spectroscopic coverage.  Although the central
values of the parameters are increased in some cases and decreased in
others, they are within error in all cases.

\subsubsection{The PSS0808+5215 field}\label{spike}

\begin{figure}
\begin{center}                          
\scalebox{0.6}[0.64]{\includegraphics{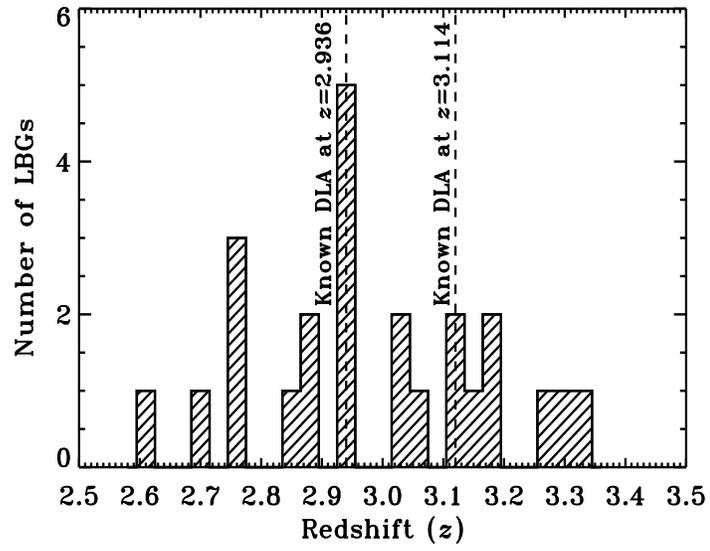}}
\caption{Redshift histogram of the $2.6<z<3.4$ LBGs in the
  PSS0808+5215 field.  Objects are binned with $\Delta z=0.03$.  The
  redshifts of the two known DLAs are indicated by the vertical dashed
  lines.  }
\label{fig:0808hist}
\end{center}
\end{figure}

Field-by-field analysis revealed a relative spike of five LBGs with
$\Delta z<0.015$ ($<10h^{-1}$Mpc) to the $z=2.936$ DLA in the
PSS0808+5215 field. A redshift histogram of the $2.6<z<3.4$ LBGs in
the PSS0808+5215 field is shown in Figure~\ref{fig:0808hist}.  To
estimate the probability of this overdensity occurring by chance, we
ran 10,000 random simulations of the distribution of the $2.6<z<3.4$
LBGs detected in the field corrected by the photometric selection
function.  We found five LBGs within $\Delta z=0.015$ of the $z=2.936$
DLA $0.16\%$ of the time.  Therefore, we conclude that this is most
likely a real overdensity.  To illustrate the extent of the
overdensity, Table~\ref{table:0808cells} lists the number of LBGs in
cells centered on the $z=2.936$ DLA with varying radius in redshift.
We find the next nearest LBG at $\Delta z=0.043$, or $29.9 h^{-1}$
Mpc.

\begin{figure}
\begin{center}                          
\scalebox{0.7}[0.7]{\includegraphics{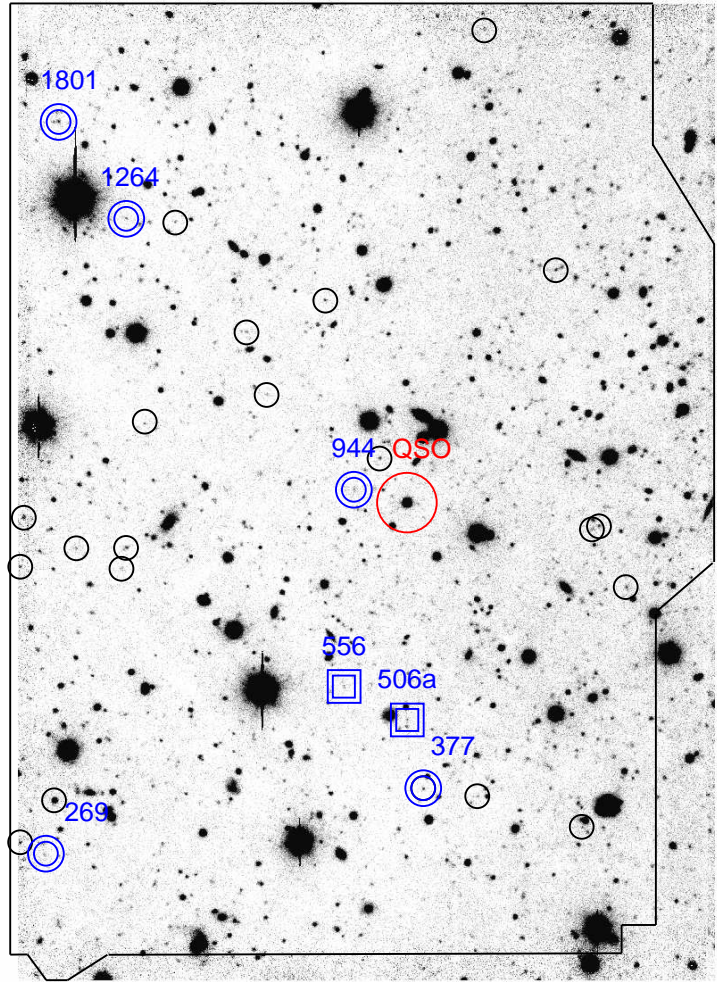}}
\caption{R-band image of LBGs identified in the PSS0808+5215 field.
  The central $z=4.45$ QSO displays DLAs at $z=2.936$ and $z=3.114$.
  The QSO is indicated by the large (red) circle.  The five LBGs
  associated with the $z=2.936$ DLA are shown using two concentric
  (blue) circles and the two LBGs associated with the $z=3.114$ DLA
  are shown using two concentric (blue) squares.  All other
  spectroscopically confirmed LBGs are shown as small (black) circles.
  The LRIS field size is $\sim5.5'\times 7'$ which corresponds to
  comoving $\sim7 \times 10 h^{-1}$Mpc.  The region bounded in black
  is the area covered by the three slitmasks used on this field. }
\label{fig:0808image}
\end{center}
\end{figure}

Inspection of the angular distribution of the LBGs in the
two-dimensional image indicates we may not be seeing the full extent
of the overdensity because of the relatively small field of view of
the LRIS camera.  In fact, this is true for all fields imaged with the
LRIS camera.  Figure~\ref{fig:0808image} presents an R-band image of
the PSS0808+5215 field.  The QSO and spectroscopically confirmed $z>2$
LBGs are marked in the image with the LBGs near the $z=2.936$ and
$z=3.114$ DLA are indicated separately.  The DLA at $z=2.936$ appears
to reside toward the apparent edge of the overdensity.  This is an
excellent argument for the acquisition of wide-field images in future
surveys and is one of the main objectives of our One-Degree Deep
survey.  Clustering analysis on scales much larger than the
correlation length is necessary for the proper correlation analysis
and study of large-scale behavior of LBGs and DLAs.

To test how this overdensity affected the overall DLA-LBG clustering
amplitude, we computed the strength of the cross-correlation in the
absence of the $z=2.936$ DLA.  We found best fit values and $1 \sigma$
errors of $r_0 = 2.98\pm 1.34$ and $\gamma = 1.32\pm0.34$ using the
conventional binning technique and $r_0 = 2.72^{+1.8}_{-2.1}$, $\gamma
= 1.48^{+1.5}_{-1.1}$ using the maximum likelihood method. The
survival of the clustering signal after the omission of the $z=2.936$
DLA in the PSS0808+5215 field is further evidence for an overdensity
of LBGs near DLAs on average.  In fact, from every measurement and
test in this work, a non-zero clustering signal has been detected.

\section{Galaxy bias and mass}\label{mass} 

The primary objective of this survey was to measure the DLA-LBG
cross-correlation function to estimate the DLA galaxy bias in the
context of CDM cosmology and use this information to infer the average
halo mass of DLAs.  This provides a first step in establishing the
fundamental properties of the population of proto-galaxies that DLAs
represent.  We were successful in making an independent measurement of
the LBG auto-correlation function at $z\sim3$ and used this as an
important calibrator to measure the DLA-LBG cross-correlation
function.  Although the uncertainties in this work make a direct
measure of the DLA bias difficult, it can be estimated in the
following way.  The relationship between the LBG auto-correlation
function $\xi_{LBG}$ and dark matter correlation function $\xi_{DM}$
on scales where the linear bias is a good model is
\begin{equation}
\xi_{LBG}(r) = b_{LBG}^2 ~\xi_{DM}(r),
\end{equation}
\noindent where $b_{LBG}$ is the LBG galaxy bias.  Similarly, for the
DLA-LBG cross-correlation function the relationship is 
\begin{equation}
\xi_{DLA-LBG}(r) = b_{DLA}b_{LBG} ~\xi_{DM}(r)
\end{equation}
\noindent \citep{eg01} where $b_{DLA}$ is the DLA galaxy bias.
Therefore, the ratio of the two relationships becomes
\begin{equation}
\frac{\xi_{DLA-LBG}(r)}{\xi_{LBG}(r)} = \frac{b_{DLA}}{b_{LBG}}.
\end{equation}

\noindent Assuming, as we have throughout this paper, that $\xi(r)$ is
well fit by a power law of the form $\xi(r) = (r/r_0)^{-\gamma}$, and
assuming identical values of $\gamma$ for both the auto-correlation
and cross-correlation functions, this ratio becomes
\begin{equation}
\left(\frac{r_{0 LBG}}{r_{0 DLA-LBG}}\right)^{-\gamma} =
\frac{b_{DLA}}{b_{LBG}}
\end{equation}

\noindent and illustrates that the ratio of the correlation lengths is
a direct indicator of the ratio of the biases.  These assumptions are
reasonable, especially when considering that both $r_0$ and $\gamma$
were freely varied when fitting the LBG auto-correlation and DLA-LBG
cross-correlation functions using each technique and produced
consistent values within their uncertainties
(Tables~\ref{table:LLresults} \&~\ref{table:DLresults}).  Moreover, we
measured the best fit values of $r_0$ for each correlation function at
various values of fixed $\gamma$ and found all resulting correlation
lengths to be in agreement within error as well.
Table~\ref{table:fixedgamma} displays the best fit values for a fixed
value of $\gamma=1.6$.

It is true that the DLA-LBG cross-correlation function measurement
from each method individually can only confirm a non-zero DLA galaxy
bias with $\sim1-2 \sigma$ confidence.  But the implications from the
combined set of measurements and tests of those measurements are what
drive our overall claim that DLAs and LBGs likely have a similar
spatial distributions and galaxy bias.  The results indicate not only
an overdensity of LBGs near DLAs over random, but correlation
functions of similar form and strength.  We find that the average
correlation lengths and uncertainties for fixed and varied values of
$\gamma$ for the LBGs in this work correspond to an average $z\sim3$
LBG galaxy bias between $2<b_{LBG}<3$ and $1.5<b_{LBG}<3$
respectively.  Similarly, the average $z\sim3$ DLA galaxy bias ranges
between $1.3<b_{DLA}<4$ and $1.5<b_{DLA}<4$ for the 11 DLAs in this
survey and $1.3<b_{DLA}<3$ and $0.8<b_{DLA}<3$ for the combined set of
15 DLAs.  The average halo mass of a galaxy population can be inferred
from the galaxy bias using halo mass function approximations, e.g.,
\cite{m98,st01}.  The above galaxy bias values correspond to LBG mass
ranges of approximately $10^{10.8}< \langle M_{LBG}\rangle<
10^{11.6}~M_{\odot}$ and $10^{9.7}< \langle M_{LBG}\rangle<
10^{11.6}~M_{\odot}$, respectively.  The average measurements for the
11 DLAs in this survey lead to approximate mass ranges of $10^{9}<
\langle M_{LBG}\rangle< 10^{11.9}~M_{\odot}$ and $10^{9.7}< \langle
M_{LBG}\rangle< 10^{12}~M_{\odot}$ and approximate mass ranges for the
combined set of 15 DLAs of $10^{9}< \langle M_{LBG}\rangle<
10^{11.6}~M_{\odot}$ and $10^{7.3}< \langle M_{LBG}\rangle<
10^{11.7}~M_{\odot}$, for fixed and varied values of $\gamma$
respectively in each case.  Both the galaxy bias and mass calculations
were determined by the method outlined in \citet{q06}.

The $z\sim3$ LBG correlation length computed by A03 is $3.96\pm0.29$
for the $\cal{R}$ $<25.5$ spectroscopic sample results in an LBG
galaxy bias of $b_{LBG}\sim4$ and corresponds to an average halo mass
of $\langle M_{LBG}\rangle\sim 10^{11.9}~M_{\odot}$.  In addition, it
has been shown that the LBG correlation length is dependent on the
observed $\cal{R}$-band (rest-frame $\sim1700$\AA) luminosity.
\citet{giav01} find average $z\sim3$ LBG masses of $\langle
M_{LBG}\rangle\sim 10^{12.4}, 10^{12},$ and $10^{11.6}~M_{\odot}$ for
LBGs with luminosities of $\cal{R}$ $=23, 25.5,$ and
$\cal{R}$$_{equiv}=27.0$, respectively, from ground-based and
space-based images [also see \cite{foucaud03,a05}].  Similarly at
$z\sim4$, \citet{kashi05} find estimated halo masses of $\langle
M_{LBG}\rangle\sim 10^{11.7-12}~M_{\odot}$ for $23.5<i'<25.5$ LBGs,
$\langle M_{LBG}\rangle\sim 10^{11.5-11.7}~M_{\odot}$ for
$25.5<i'<26.5$ LBGs, and $\langle M_{LBG}\rangle\sim
10^{11.3-11.5}~M_{\odot}$ for $26.5<i'<27.4$ LBGs.  From these
relationships, it can be assumed that the typical mass of LBGs is
below that of the $\cal{R}$ $<25.5$ (and R $<25.5$) spectroscopic
sample.

\section{Discussion}\label{disc} 

In addition to the agreement between the clustering behavior and
implied masses of DLAs and LBGs from this work, there appears to be
mounting evidence in favor of the idea that high redshift DLAs and
LBGs sample the same population [e.g., \citet{joop01}] such as: (1)
The two-dimensional $z\sim3$ DLA-LBG cross-correlation analysis of
\citet{bl04} of two DLAs and one sub-DLA in wide-field images found a
non-zero clustering amplitude to more than $2 \sigma$ using a profile
similar to the LBG auto-correlation function and sampling the behavior
of DLAs with LBGs on angular scales equal to and beyond those in this
work ($\sim1-15h^{-1}$Mpc), (2) Two $z>2$ DLAs detected in emission
and examined in Hubble Space Telescope ({\it HST}) images exhibit
properties consistent with those of the LBG population \citep{m02},
(3) The $z\sim3$ DLA heating rates implied by the \textsc{Cii*} method
\citep{w03} require localized nearby sources of star formation that
are consistent with those found for average LBGs, (4) The Lyman
$\alpha$ emission of a $z=2.04$ DLA detected in the trough of the
Lyman $\alpha$ absorption feature in the spectrum of PKS 0458-02
\citep{m04} is consistent with LBG Lyman $\alpha$ emission, (5) The
dearth of faint high redshift sources having low {\it in situ} star
formation rates that meet the criteria required by DLA statistics
\citep{chen06} in the {\it HST} UDF images, (6) The DLA-LBG
correlation length of $r_0\lesssim 2.85h^{-1}$Mpc determined by the
hydrodynamic simulations of \citet{bl05} is in very good agreement
with the results from this work and is consistent with the LBG
population as a whole, (7) The results from high resolution numerical
simulations of \citet{nag04a,nag04b,nag06} indicate that strong
galactic-scale winds from starbursts evacuate the gas in lower-mass
$(\langle M_{DLA}\rangle \lesssim 10^8 M_{\odot})$ DLAs driving up the
mean DLA mass in the stronger galactic-wind scenarios to values in
good agreement with this work and average LBGs, and (8) The
typical LBG magnitude of R $\sim27$ and small impact parameter of
$<1''$ is consistent with the very few detections of DLA emission in
the sight-lines to QSOs.

One picture that could reconcile the above results is where LBGs are
starbursting regions embedded in relatively massive DLA systems.  The
LBG starbursts can provide the necessary heating to explain the
observed \textsc{Cii*} observations under reasonable geometric
assumptions.  Although signal-to-noise ratio of the low-resolution
spectra of individual LBGs is poor, the Lyman $\alpha$ features that
are observed in a subset of LBG spectra appear to be damped. This
includes the high signal-to-noise ratio exception of the
gravitationally lensed LBG MS1512-cB58 \citep{pet02}. The dynamics of
large systems are able to explain the observed DLA gas kinematics as
shown in \citet{pw97,pw98,wp00a,wp00b}. The large radial distribution
of cold gas necessary in the semi-analytical models of
\citet{maller01} support this picture, however, large systems with
these properties at high redshift are sometimes difficult to rectify
in popular models.  In addition, the average metallicity of LBGs
($Z/Z_{\odot}\sim1/4$) is typically higher than that of DLAs
($Z/Z_{\odot}\sim1/30$), yet the lowest LBG metallicities and highest
DLA metallicities overlap.  Any perceived discrepancy is likely to be
resolved by invoking metallicity gradients and applying feedback,
dust, and multi-phase arguments, such as those proposed in
\citet{nag04b}, to future simulations.  What is needed observationally
are ground-based surveys focused on the spatial distribution of DLAs
and LBGs in addition to space-based and AO observations revealing the
luminosity function and morphologies of DLAs.

\section{Summary}\label{resultssummary}   

Our survey for galaxies associated with DLAs at $z\sim3$ has been
successful in developing an efficient $u'$BVRI photometric selection
technique and color criteria to detect LBGs in QSO fields with known
DLAs. We used 211 $z>2$ LBG spectra to make an independent measurement
of the three-dimensional LBG auto-correlation function and the first
measurement of the three-dimensional DLA-LBG cross-correlation
function.

\begin{figure}
\begin{center}                         
\scalebox{0.32}[0.28]{\rotatebox{90}{\includegraphics{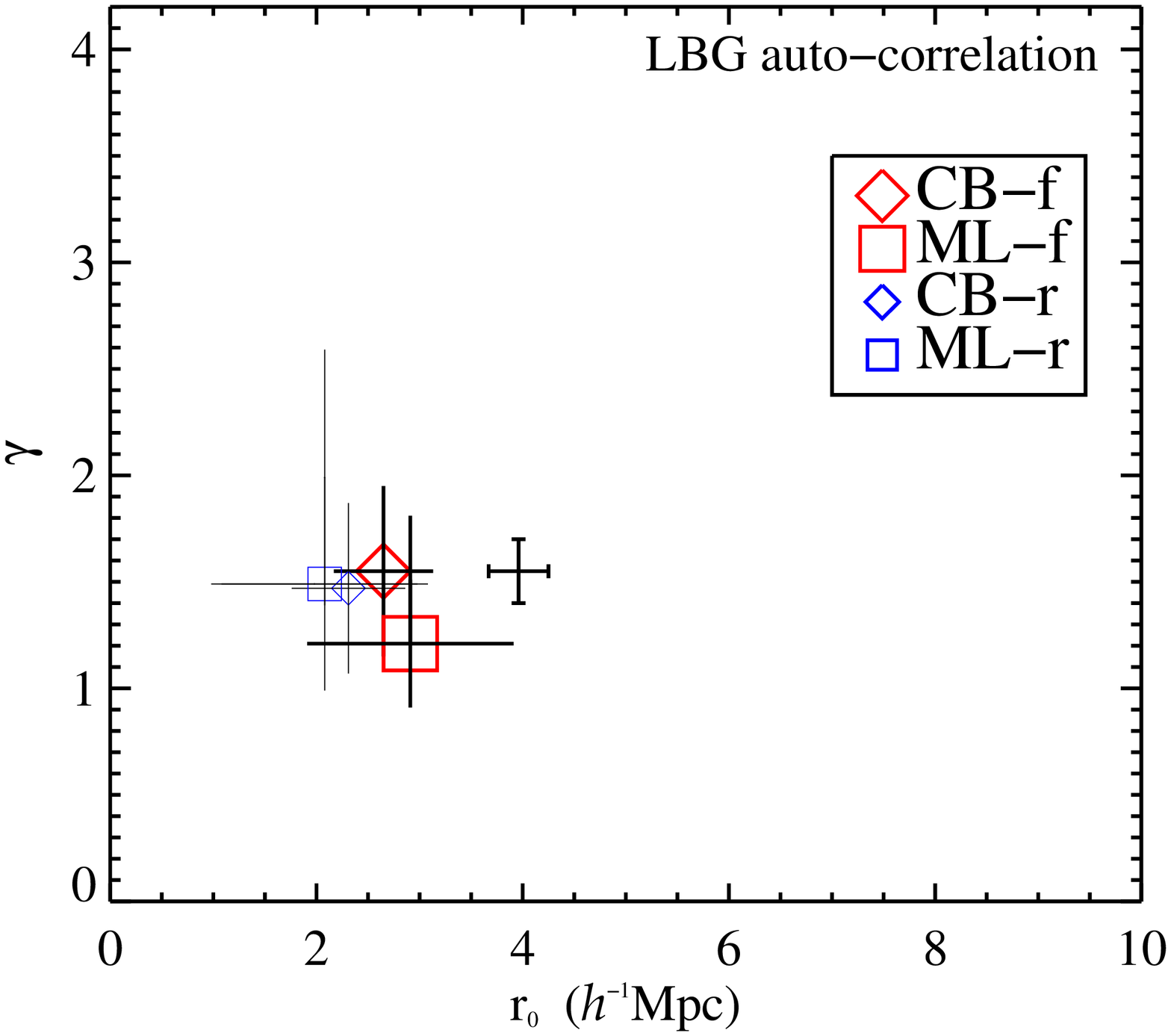}}}
\end{center}
\begin{center}                         
\scalebox{0.32}[0.28]{\rotatebox{90}{\includegraphics{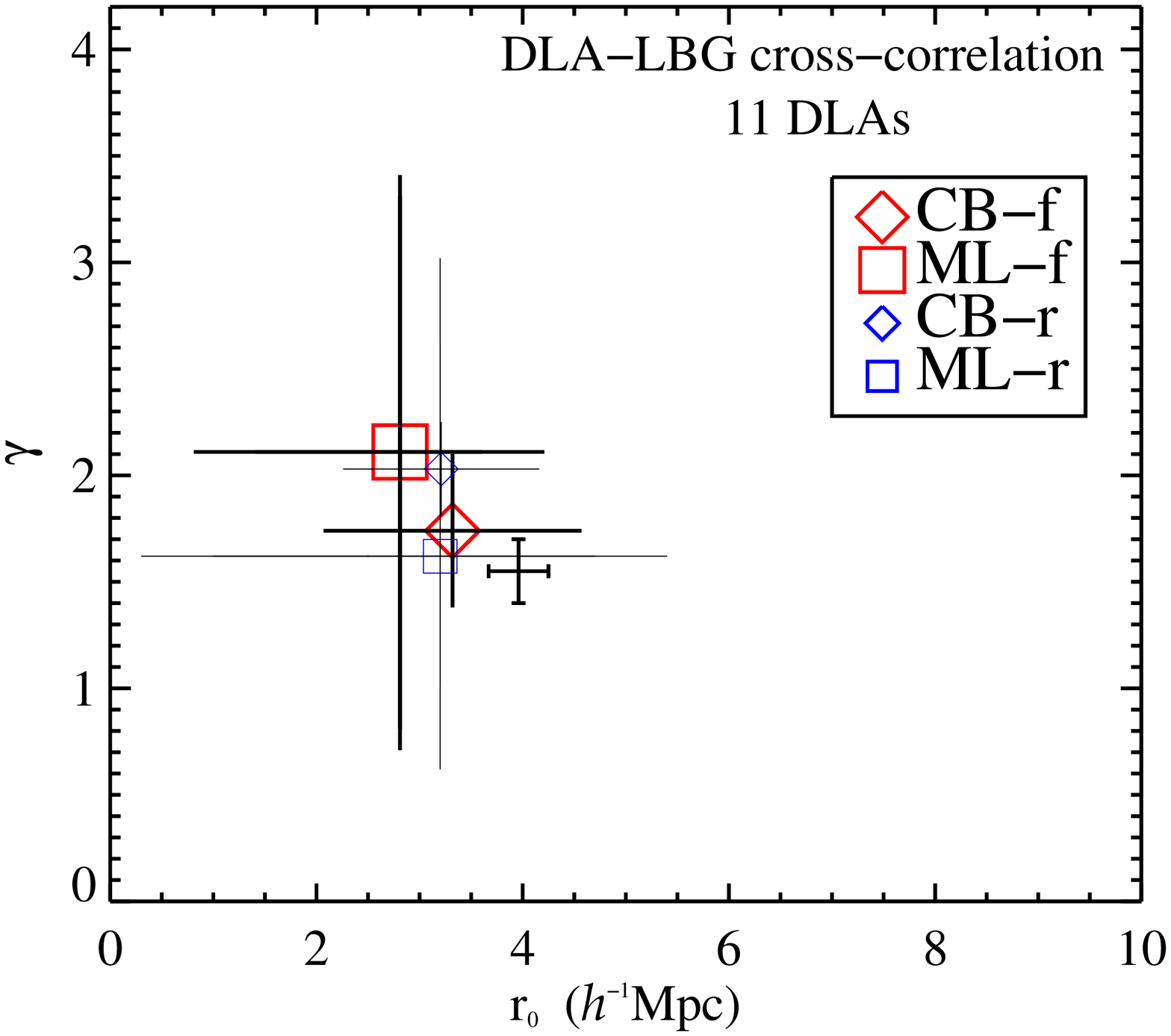}}}
\end{center}
\begin{center}                         
\scalebox{0.32}[0.28]{\rotatebox{90}{\includegraphics{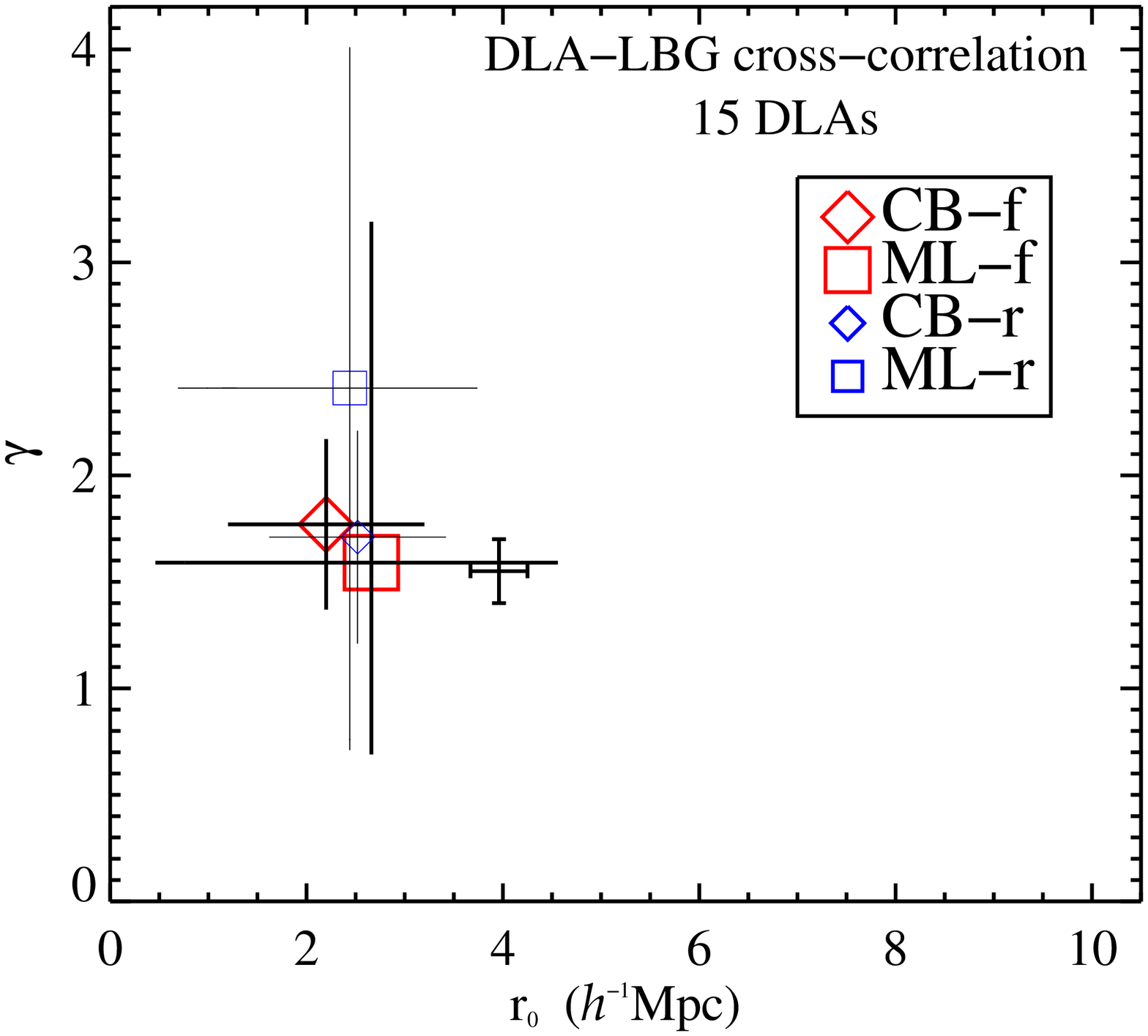}}}
\caption
{\small Results summary plots for the LBG auto-correlation and DLA-LBG
  cross-correlation functions.  The best fit values and $1\sigma$
  uncertainties are shown for the LBG auto-correlation (upper panel),
  the DLA-LBG cross-correlation for the 11 DLAs in this survey (center
  panel), and the full set of 15 DLAs (lower panel) that includes 4
  DLAs from the available dataset of \citet{s03}.  Large bold symbols
  indicate the results for the correlation function measurements and
  small symbols indicate the results for the tests on these
  measurements.  The values for the LBG auto-correlation determined by
  \citet{a03} are shown by the error cross at $r_0=3.96,\gamma=1.55$
  for comparison.  The legend code is as follows: CB = Conventional
  binning, ML = Maximum likelihood, f = galaxies in the random
  catalogs have the fixed angular positions of the data, r = galaxies
  in the random catalogs have random positions.}
\label{fig:resultsplot}
\end{center}
\end{figure}

We used a modified version of the conventional binning technique
following the prescription in A03 and measured best fit values and $1
\sigma$ uncertainties of $r_0=2.65\pm 0.48$ and $\gamma=1.55\pm 0.40$
for the $z\sim3$ LBG auto-correlation function.  These results are in
agreement with the previous measurement by A03 when considering that
the uncertainties may be underestimated by a factor of $1-2$
(\S~\ref{adel_scheme}).  Applying this technique to the DLA-LBG
cross-correlation resulted in best fit values and $1 \sigma$ errors of
$r_0=3.32\pm 1.25$ and $\gamma=1.74\pm 0.36$ for the set of 11 DLAs in
our survey and $r_0=2.20\pm0.96$, $\gamma=1.73\pm0.39$ for the
combined set of 15 DLAs that include 4 DLAs from the survey of S03.
These results are shown as large (red) diamonds in
Figure~\ref{fig:resultsplot}.

Although the above binning technique can produce accurate results,
conventional binning techniques in general are dependent on bin size,
interval, and bin center.  To get around these dependencies, we
independently measured the correlation functions using a maximum
likelihood technique based on Poisson statistics.  This method is
bin-independent, makes full use of the data, and is ideal for small
datasets.  We found maximum likelihood values and $1\sigma$ confidence
levels of $r_0 = 2.91^{+1.0}_{-1.0}, \gamma = 1.21^{+0.6}_{-0.3}$ for
the LBG auto-correlation function and $r_0 = 2.81^{+1.4}_{-2.0},
\gamma = 2.11^{+1.3}_{-1.4}$ and $r_0 = 2.66^{+1.9}_{-2.1}, \gamma =
1.59^{+1.6}_{-0.9}$ for the DLA-LBG cross-correlation functions for
the sets of 11 and 15 DLAs, respectively.  The results for both the
auto-correlation and the cross-correlation are indicated by large
(red) squares in Figure~\ref{fig:resultsplot}.

We tested the effects that the physical constraints of the slitmasks
have on the angular component of the correlation function by
re-analyzing the data by means of the above two techniques and
assigning random angular positions to the random galaxy catalogs
instead of the angular positions of the data as was imposed in the
original analysis. The test results using the conventional binning
technique are shown as small (blue) diamonds in
Figure~\ref{fig:resultsplot} and as small (blue) squares using the
maximum likelihood technique.

Furthermore, we discovered a relative spike of five LBGs within
$\Delta z=0.015$ of the $z=2.936$ DLA in the field PSS0808+5215 and
tested the average DLA-LBG clustering signal in the absence of this
DLA.  As expected, the amplitude of the clustering was diminished, but
the overdensity and form of the correlation function survives and
remains in good agreement with the values determined for the complete
set of DLAs.

Lastly, we found that two of the six QSOs in the survey spectra lie
within $\Delta z=0.0125$ of two of the DLAs. We determine a 1 in 940
chance of this occurring randomly and interpret this to suggest a
possible relationship between the distribution of QSOs and DLAs at
$z\sim3$.  If found to be a common occurence, the close proximity of
QSOs to DLAs could lend insight into the duty-cycle of QSOs and the
size and persistence of systems with high \textsc{Hi} column densities
near sources of significant ionizing radiation.

It can be seen from Figure~\ref{fig:resultsplot} that all of the
independent methods varying both $r_0$ and $\gamma$, and tests of
those methods, produce results that are in agreement within their
uncertainties.  This is also true for the measurements of $r_0$
holding $\gamma$ fixed when using all methods
(Table~\ref{table:fixedgamma}).  The DLA-LBG cross-correlation
function, determined from both the set of 11 DLAs in our survey and
the combined set of 15 DLAs, exhibits a measurable clustering signal
and has best fit parameters in agreement with those of the LBG
auto-correlation function. The individual measurements of the DLA-LBG
cross-correlation function are only able to measure the DLA galaxy
bias to $1-2 \sigma$ and we expect to improve this measurement to
$\sim3 \sigma$ significance when combining these results with those of
our One-Degree Deep survey in progress.  When letting $r_0$ and
$\gamma$ vary, we found that the best fit values and their
uncertainties suggest an LBG galaxy bias of $1.5<b_{LBG}<3$
corresponding to an average halo mass of $10^{9.7}< \langle
M_{LBG}\rangle< 10^{11.6}~M_{\odot}$.  The DLA galaxy bias determined
identically for the 11 DLAs in this survey is $1.3<b_{DLA}<4$ and
infers an average DLA mass of $10^{9.7}< \langle M_{LBG}\rangle<
10^{12}~M_{\odot}$.  These values are $0.8<b_{DLA}<4$ and $10^{7.3}<
\langle M_{LBG}\rangle< 10^{11.7}~M_{\odot}$ for the combined set of
15 DLAs.  Lastly, we offer the plausible scenario that LBGs reside in
the same systems that host DLAs.  We identify several pieces of
evidence in the literature that support this view.  Whatever the true
picture, the results from this survey have shed light on the elusive
mass of DLAs, their distribution with LBGs, and a possible link
between the distribution of QSOs and DLAs at $z\sim3$.
 
\acknowledgments

The authors wish to thank C. Mullis and K. L. Adelberger for very
helpful discussions. In addition, we would like to thank A. Dressler,
R. Giovanelli, and J. Darling for access to the Palomar Observatory
and observational assistance with the COSMIC camera.  The authors
recognize and acknowledge the very significant cultural role and
reverence that the summit of Mauna Kea has always had within the
indigenous Hawaiian community.  We are most fortunate to have the
opportunity to conduct observations from this mountain.  This work was
partially supported by the National Science Foundation grant
AST-0307824 and the NSF Astronomy \& Astrophysics Postdoctoral
Fellowship (AAPF) grant AST-0201667 awarded to Eric Gawiser.



\clearpage
\begin{landscape}
\begin{deluxetable}{lcccccc}
\tabletypesize{\normalsize}
\tablecaption{Survey Fields \label{table:qsofields}}
\tablewidth{0pt}
\tablehead{
\colhead{Field} & \colhead{R.A.(J2000.0)} & \colhead{Dec.(J2000.0)} &
\colhead{$z_{QSO}$} & \colhead{$z_{DLA(s)}$} & 
\colhead{log N(\textsc{Hi})} & \colhead{$E(B-V)$}\\ 
\colhead{} & \colhead{$(h~~m~~s)$} & \colhead{$(d~~m~~s)$} 
& \colhead{} & \colhead{} & \colhead{} & \colhead{}}
\startdata
LBQS0056+0125 & 00 59 17.62 & +01 42 05.30 & 3.149 & 2.775 
 & $21.0$\tablenotemark{1} &  0.03\\
PKS0336--017  & 03 39 00.65 & $-$01 33 19.20 & 3.197 & 3.062 
 & $21.2$\tablenotemark{2} &  0.14\\ 
PSS0808+5215  & 08 08 49.43 & +52 15 14.90 & 4.450 & 2.936,~3.113 
 & $20.9$\tablenotemark{3},~$20.7$\tablenotemark{4} &  0.04\\
PSS0957+3308  & 09 57 44.50 & +33 08 23.00 & 4.250 & 3.280
 & $20.5$\tablenotemark{4} &  0.01\\
BRI1013+0035  & 10 15 48.96 & +00 20 19.52 & 4.381 & 3.103 
 & $21.1$\tablenotemark{5} &  0.03\\
PSS1057+4555  & 10 57 56.39 & +45 55 51.97 & 4.116 & 3.050,~3.317 
 & $20.3$\tablenotemark{5}~$^{,}$\tablenotemark{6}~, 
$20.3$\tablenotemark{7} &  0.01\\
PSS1432+3940  & 14 32 24.90 & +39 40 24.00 & 4.280 & 3.272 
 & $21.3$\tablenotemark{4} &  0.01\\
PC1643+4631A  & 16 45 01.09 & +46 26 16.44 & 3.790 & 3.137 
 & 20.7\tablenotemark{8} &  0.02\\
JVAS2344+3433 & 23 44 51.25 & +34 33 48.64 & 3.053 & 2.908 
 & $21.1$\tablenotemark{4} & 0.08\\
\tablenotetext{1}{\citet{w95}}
\tablenotetext{2}{\citet{p01}}
\tablenotetext{3}{\citet{c05}}
\tablenotetext{4}{\citet{p03}}
\tablenotetext{5}{\citet{slw00}}
\tablenotetext{6}{\citet{celine01}}
\tablenotetext{7}{\citet{lu98}}
\tablenotetext{8}{\citet{sch91}}
\enddata
\end{deluxetable}
\end{landscape}

\clearpage
\begin{deluxetable}{ccc}
\tabletypesize{\normalsize} 
\tablecaption{PSS0808+5215 Cell Counts
\label{table:0808cells}} 
\tablewidth{0pt} \tablehead{ \colhead{$r_z$} & 
\colhead{$h^{-1}$ Mpc\tablenotemark{a}} & 
\colhead{LBGs\tablenotemark{b}}} 
\startdata
0.0025 &  1.68 & 2 \\ 
0.0050 &  3.36 & 2 \\ 
0.0075 &  5.03 & 3 \\ 
0.0100 &  6.71 & 3 \\ 
0.0125 &  8.38 & 3 \\ 
0.0150 & 10.05 & 5 \\
\vdots & \vdots & 5 \\ 0.0450 & 29.99 & ~~6
\tablenotetext{a}{$\Omega_M$=0.3, $\Omega_\Lambda$=0.7}
\tablenotetext{b}{Number of LBGs found in cells with dimensions
  $\sim5.5'\times 7'$ ($\sim7\times10~h^{-2}$Mpc$^2$ at $z=3$) and
  $2\cdot r_z$ centered on the $z=2.936$ DLA.  Typical errors in LBG
  redshifts caused by galactic-scale winds are $\sim2h^{-1}$Mpc.}
\enddata
\end{deluxetable}

\clearpage
\begin{deluxetable}{lcc}
\tablecaption{LBG Auto-Correlation Parameter Summary
\label{table:LLresults}}
\tablewidth{0pt}
\tablehead{
\colhead{Method} & \colhead{$r_0$} & \colhead{$\gamma$}}
\startdata
Conventional Binning\tablenotemark{a}~$^{,}$\tablenotemark{b} &
              $2.65\pm0.5$ & $1.55\pm0.4$ \\
Maximum Likelihood\tablenotemark{b} &
              $2.91^{+1.0}_{-1.0}$ & $1.21^{+0.6}_{-0.3}$ \\
\hline
\underline{Tests} & & \\
Conventional Binning\tablenotemark{a}~$^{,}$\tablenotemark{c} & 
               $2.31\pm0.6$ & $1.47\pm0.4$ \\
Maximum Likelihood\tablenotemark{c}  &    
               $2.08^{+1.0}_{-1.1}$ & $1.49^{+1.1}_{-0.5}$ \\ 
\enddata
\tablenotetext{a}{Galaxy separations determined using a cylindrical
  approach described in \citet{a03}, Appendix C}
\tablenotetext{b}{Angular positions of galaxies in the random
  catalogs are identical to the angular positions of the data}
\tablenotetext{c}{Angular positions of galaxies in the random 
  catalogs are random}
\end{deluxetable}

\begin{deluxetable}{lcc}
\tablecaption{DLA-LBG Cross-Correlation Parameter Summary
\label{table:DLresults}}
\tablewidth{5.45in} 
\tablehead{ 
\colhead{Method} & \colhead{11 DLAs\tablenotemark{a}} & 
\colhead{15 DLAs\tablenotemark{b}}\\
\colhead{} & \colhead{$r_0$~~~~~~~~~~~~~~~~~~$\gamma$} &
\colhead{$r_0$~~~~~~~~~~~~~~~~~~$\gamma$}} 
\startdata Conventional Binning\tablenotemark{c}~$^{,}$\tablenotemark{d} 
 & $3.32\pm1.3$ ~~$1.74\pm0.4$ & $2.20\pm1.0$ ~~ $1.77\pm0.4$\\ 
Maximum Likelihood\tablenotemark{d}
 & $2.81^{+1.4}_{-2.0}$ ~~~~~~$2.11^{+1.3}_{-1.4}$ & $2.66^{+1.9}_{-2.1}$ 
~~~~~~$1.59^{+1.6}_{-0.9}$\\ 
\hline \underline{Tests} & & \\ 
Conventional Binning\tablenotemark{c}~$^{,}$\tablenotemark{e}
 & $3.21\pm1.0$ ~~$2.03\pm0.2$ & $2.52\pm0.9$ ~~ $1.71\pm0.5$\\ 
Maximum Likelihood\tablenotemark{e} & $3.20^{+2.2}_{-2.9}$ ~~~~~~
$1.62^{+1.4}_{-1.0}$ & $2.44^{+1.3}_{-1.9}$ ~~~~~~$2.41^{+1.6}_{-1.7}$\\ 
\enddata 
\tablenotetext{a}{Results using the 11 DLAs from this work} 
\tablenotetext{b}{Results from this work combined with the DLA and LBG 
  information for the four DLAs in the survey of
\citet{s03}} 
\tablenotetext{c}{Galaxy separations determined using a
  cylindrical approach described in \citet{a03}, Appendix C}
\tablenotetext{d}{Angular positions of galaxies in the random catalogs
  are identical to the angular positions of the data}
\tablenotetext{e}{Angular positions of galaxies in the random catalogs
  are random}
\end{deluxetable}

\begin{deluxetable}{lccc}
\tablecaption{Correlation Lengths for $\gamma=1.6$
\label{table:fixedgamma}}
\tablewidth{0pt}
\tablehead{
\colhead{Method} & \colhead{LBG-LBG} & 
\colhead{DLA-LBG(11)\tablenotemark{a}} & 
\colhead{DLA-LBG(15)\tablenotemark{b}}}
\startdata
Conventional Binning\tablenotemark{c}~$^{,}$\tablenotemark{d}
& $2.72\pm0.5$ & $3.53\pm1.0$ & $2.27\pm1.0$\\ 
Maximum Likelihood\tablenotemark{d}
& $3.32^{+0.6}_{-0.6}$ & $2.93^{+1.4}_{-1.5}$ & $2.66^{+1.2}_{-1.3}$\\ 
\hline
\underline{Tests} & & \\
Conventional Binning\tablenotemark{c}~$^{,}$\tablenotemark{e}
& $2.40\pm0.5$ & $3.87\pm1.0$ & $2.53\pm0.8$\\  
Maximum Likelihood\tablenotemark{e} 
& $2.96^{+0.6}_{-0.7}$ & $3.20^{+1.5}_{-1.7}$ & $2.61^{+1.3}_{-1.5}$\\ 
\enddata
\tablenotetext{a}{Results using the 11 DLAs from this work}
\tablenotetext{b}{Results from this work combined with the DLA and LBG
  information for the four DLAs in the survey of \citet{s03}}
\tablenotetext{c}{Galaxy separations determined using a cylindrical
  approach described in \citet{a03}, Appendix C}
\tablenotetext{d}{Angular positions of galaxies in the random
  catalogs are identical to the angular positions of the data}
\tablenotetext{e}{Angular positions of galaxies in the random 
  catalogs are random}
\end{deluxetable}


\clearpage

\clearpage

\clearpage

\clearpage

\clearpage

\clearpage

\clearpage

\clearpage

\clearpage

\clearpage

\clearpage

\clearpage

\end{document}